\documentclass[fleqn,usenatbib]{mnras}

\usepackage{newtxtext,newtxmath}

\usepackage[T1]{fontenc}
\usepackage{ae,aecompl}

\usepackage{graphicx}	
\usepackage{amsmath}	
\usepackage{hyperref}
\usepackage{enumerate}
\usepackage{multirow}
\usepackage{multicol}

\title[Timescales of Interaction]{Timescales for the Effects of Interactions on Galaxy Properties and SMBH Growth}

\author[D. O'Ryan et al.]{David O'Ryan$^{1,2,3}$\thanks{E-mail: \href{mailto:david.oryan@esa.int}{david.oryan@esa.int}},
    Brooke D. Simmons$^{1}$,
    Andreas L. Faisst$^{4}$,
    Izzy L. Garland$^{5}$,
    \newauthor
    Tobias G\'eron$^{6}$,
    Ghassem Gozaliasl$^{7}$,
    Steven Gillman$^{8,9}$,
    Sofia Guedes Vaz Pinto$^{1}$,
    \newauthor
    William C. Keel $^{10}$,
    Anton M. Koekemoer$^{11}$,
    Sandor Kruk$^{3}$,
    Karen L. Masters$^{12}$,
    \newauthor
    Oscar Montoya C.$^{1}$,
    Mason Redden$^{1}$,
    Matthew R. Thorne$^{1}$,
    Emily R. Walls$^{1}$,
    \newauthor
    Deneth Weerasinghe$^{1}$
    \& John R. Weaver$^{13}$
\\
$^{1}$Department of Physics, Lancaster University, Bailrigg, Lancaster, LA1 4YB, UK \\
$^{2}$Centro de Astrobiologia, CSIC-INTA, ESAC Campus, 28962 Villanueva de la Canada, Madrid, Spain \\
$^{3}$European Space Agency (ESA), European Space Astronomy Centre (ESAC), Camino Bajo del Castillo s/n, 28692, Villaneuva de la Cañada, Madrid, Spain \\
$^{4}$Caltech/IPAC, 1200 E. California Blvd. Pasadena, CA 91125, USA \\
$^{5}$Department of Theoretical Physics and Astrophysics, Faculty of Science, Masaryk University, Kotl\'{a}\v{r}sk\'{a} 2, Brno, 611 37, Czech Republic \\
$^{6}$Dunlap Institute for Astronomy and Astrophysics, University of Toronto, 50 St. George Street, Toronto, ON M5S 3H4, Canada \\
$^{7}$Department of Computer Science, Aalto University, PO Box 15400, Espoo, FI-00 076, Finland \\
$^{8}$Cosmic Dawn Centre (DAWN), Denmark  \\
$^{9}$DTU Space, Technical University of Denmark, Elektrovej, Building 328, 2800, Kgs. Lyngby, Denmark \\
$^{10}$Department of Physics and Astronomy, University of Alabama, Box 870324, Tucaloosa, AL 35487 \\
$^{11}$Space Telescope Science Institute, 3700 San Martin Dr., Baltimore, MD 21218, USA  \\
$^{12}$Departments of Physics and Astronomy, Haverford College, Lancaster Avenue, Ardmore, PA 19041 USA \\
$^{13}$Department of Astronomy, University of Massachusetts, Amherst, MA 01003, USA \\
}

\date{Accepted 28/03/2025. Received 03/12/2024; in original form 03/12/2024}

\pubyear{2024}

\begin{document}
\label{firstpage}
\pagerange{\pageref{firstpage}--\pageref{lastpage}}
\maketitle

\begin{abstract}
   Galaxy interaction and merging have clear effects on the systems involved. We find an increase in the star formation rate (SFR), potential ignition of active galactic nuclei (AGN) and significant morphology changes. However, at what stage during interactions or mergers these changes begin to occur remains an open question. With a combination of machine learning and visual classification, we select a sample of 3,162 interacting and merging galaxies in the Cosmic Evolutionary Survey (COSMOS) field across a redshift range of 0.0 - 1.2. We divide this sample into four distinct stages of interaction based on their morphology, each stage representing a different phase of the dynamical timescale. We use the rich ancillary data available in COSMOS to probe the relation between interaction stage, stellar mass, SFR, and AGN fraction. We find that the distribution of SFRs rapidly change with stage for mass distributions consistent with being drawn from the same parent sample. This is driven by a decrease in the fraction of red sequence galaxies (from 17\% as close pairs to 1.4\% during merging) and an increase in the fraction of starburst galaxies (from 7\% to 32\%). We find the AGN fraction increases by a factor of 1.2 only at coalescence. We find the effects of interaction peak at the point of closest approach and coalescence of the two systems. We show that the point in time of the underlying dynamical timescale - and its related morphology - is as important to consider as its projected separation.
\end{abstract}


\section{Introduction}\label{introduction}
\noindent Galaxy interaction and merging is of fundamental importance to our current theories of galaxy formation and evolution. The leading cosmological theory, $\Lambda$ Cold Dark Matter ($\Lambda$CDM), dictates that galaxies are expected to form and assemble hierarchically \citep{1978MNRAS.183..341W, 2001MNRAS.328..726S, 2013MNRAS.436.1765M}. There is further evidence of this in the number of galaxy pairs and galaxies exhibiting obvious morphological distortion, showing that interaction and merging may be very common \citep{1977egsp.conf..401T}. It is well known that many underlying processes are triggered or enhanced due to galaxy interaction. The clearest result of interaction is the disruption of galactic morphology, the acceleration of disks to spheroidal systems and the formation of tidal features \citep{1972ApJ...178..623T, 1977ApJ...212..616T, 2005MNRAS.357..753G, 2009MNRAS.397..802H}, increases in star formation rates (SFRs) when compared to the general galaxy population \citep{1991ApJ...370L..65B, 2006ApJ...652...56B, 2014MNRAS.437.2137S, 2015ApJ...807L..16K} and, potentially, their rapid quenching \citep{2013MNRAS.430.1901H, 2023RAA....23i5026D}. There is also evidence to suggest that interaction plays a role in the ignition of active galactic nuclei (AGN) \citep{2011MNRAS.418.2043E, 2015ApJ...806..219C, 2023MNRAS.523.4164H}. Therefore, with interaction having such an effect, understanding galaxy interaction is fundamental to understanding galaxy evolution.

To map out the effect of galaxy interaction, we must understand its impact through different merger histories and dynamical timescales. We define the dynamical timescale as the time which the two (or more) galaxies have a gravitational effect on one another, and retain the signatures of those effects. We often use the crossing time of the two systems over each other to define this, but in the context of this work the dynamical time will span the full time that two interacting galaxies will have a gravitational effect upon one another. Due to the timescale of an interaction, we cannot observe a continuous interaction. Instead we piece together many observational snapshots of many systems at different epochs of the dynamical time. However, observational samples are often too small to be fully representative \citep{2000ApJ...530..660B, 2012A&A...539A..45L, 2014A&A...567A.132W}. We gain insights from works focused on simulations with dynamical models, where the entire timescale is recorded \citep[e.g.][]{1990AJ....100.1477W, 2009AJ....137.3071B, 2016MNRAS.459..720H}, and large cosmological simulations \citep[e.g][]{2008MNRAS.391.1137L, 2019MNRAS.485.1320M, 2020MNRAS.493.3716H}. These simulations find increases in star formation at the mid-point of the dynamical timescale - as the two systems flyby \citep{2008MNRAS.384..386C, 2019MNRAS.490.2139R, 2021MNRAS.503.3113M}. However, confirming this observationally remains elusive due to the lack of large sample \citep{2023ApJ...958...96R}. For a similar reason, observationally confirming the relation between interaction, AGN switch on or quenching remains elusive \citep{2011MNRAS.418.2043E, 2018PASJ...70S..37G, 2023ApJ...942..107S}. 

Therefore, the method to constrain these relationships lies in creating large samples of interacting galaxies which contain enough systems to represent the full dynamical history, various orbital geometries and pre-encounter galaxy properties. Attempts to create such a sample have been either by machine learning classification \citep{2019A&A...626A..49P, 2023A&A...669A.141S}, visual classification by citizen scientists \citep{2010MNRAS.401.1043D} or by photometric parameterisation \citep{2004AJ....128..163L, 2023MNRAS.522....1N}. However, such samples have been plagued by contamination and a loss of statistical significance when broken down into stages due to the resultant low sample sizes.

An alternate approach to directly measuring the dynamical timescale of an interaction has been to infer it from the projected separation between the two systems. The projected separation is the measured distance between the two galaxies, assuming they are at the same redshift. Early works showed that there is significant star formation enhancement (SFE) in galaxy pairs with small separations \citep{2003MNRAS.346.1189L,2008MNRAS.385.1903L, 2008AJ....135.1877E, 2022ApJ...940....4S}. A similar connection was also found with the AGN fraction, where it appeared to increase with decreasing projected separation \citep{2009MNRAS.399.2172R, 2011MNRAS.418.2043E, 2017MNRAS.465.2671G, 2021ApJ...909..124S}. However, using projected separation as a proxy for the stage overlooks which part of the dynamical history the system is occupying. For instance, if a galaxy pair is found to have small projected separation, without visually confirming the morphology, it is difficult to ascertain if the galaxies are just approaching each other or have just passed each other. Extended morphological tracers can help break this degeneracy.

In this work, we split a large sample of interacting galaxies found in \citet[][hereafter OR23]{2023ApJ...948...40O} into four specific stages based on their morphology. Each stage is designed to capture different parts of the dynamical time of an interaction. The stages range from galaxies being close pairs, to overlapping and disturbed, to disturbed and distinct, to coalescing. This follows the classification methods of developing algorithms for staging of interaction \citep[e.g][]{2019MNRAS.490.5390B,2022ApJ...937...97C}. 

We focus here on the subset of the OR23 interacting galaxy sample in the Cosmic Evolutionary Survey (COSMOS) \footnote{DOI: \href{https://www.ipac.caltech.edu/doi/irsa/10.26131/IRSA178}{10.26131/IRSA178}} \citep{2007ApJS..172....1S}, specifically the Classic COSMOS2020 catalogue \citep[C20;][]{2022ApJS..258...11W}. This provides us with ancillary data which contains many galactic parameters of interest, including galaxy stellar masses, SFRs, photometric redshifts, etc. We investigate the evolution of the stellar masses, SFRs and AGN fraction at different points of the dynamical timescale. We use the photometric redshifts available to confirm a set of interacting galaxies and physically close pairs to draw on trends with projected separation.

This paper is laid out as follows in Section \ref{data} we briefly summarise our interacting galaxy catalogue and the C20 catalogue. We further describe the process of catalogue matching especially focused on de-duplication and accurate cross-matching. Section \ref{method} describes how we split the interacting systems into stages, define our interaction stage classifications, and how we identify the secondary galaxies in paired systems. In Sections \ref{results:SF_stage} and \ref{results:AGN_stage} we show our results of SFR and AGN fraction evolution with interaction stage and present an initial discussion. These sections are followed by Section \ref{discussion} where we compare our results to previous works and put them in the context of the field. Finally, in Section \ref{conclusion} we make concluding remarks and discuss future work.

Where necessary, we use a Flat $\Lambda$CDM cosmology with $H_0$ = $70$\,km/s/Mpc and $\Omega_M = 0.3$. Hereafter in this paper, when referring to an interacting galaxy we are referring to a galaxy which has undergone one or multiple flybys by a secondary galaxy and caused tidal disturbance. A merging galaxy is the final state of these flybys, where two or more systems have coalesced to form a highly morphologically irregular system.

\section{Data: Catalogue Matching \& Secondary Identification} \label{data}
\subsection{The OR23 Catalogue}\label{sec:or_cat}
\noindent We briefly describe the OR23 catalogue, how interacting galaxies were identified. The OR23 catalogue was constructed by applying a convolutional neural network - specifically the model \texttt{Zoobot} - to the entire \textit{Hubble} Source Catalogue \citep[HSC;][]{2016AJ....151..134W}. This contains approximately 92 million extended sources across a large volume of \textit{HST} observations.

Due to the data volume that would be required for this application, the platform ESA Datalabs\footnote{ESA Datalabs: \url{https://datalabs.esa.int/}} was used. ESA Datalabs allows direct access to all archival \textit{HST} observations, and avoids the need to download any observational data.

The 92 million source cutouts were classified into two classes: interacting and non-interacting galaxies. Each cutout was a single band image, created from observations with the $F814W$ filter and only from \textit{Hubble} Advanced Product (HAP) data. Each cutout was fixed at 150 $\times$ 150 pixels, and was created using a \texttt{ZScaleInterval} and a \texttt{LinearStretch} in the Astropy Python package \citep{astropy:2022}. Upon classification, de-duplication of shredded galaxies and manual removal of contaminants left a final candidate catalogue of 21,926 interacting galaxies. 

This catalogue, however, was from a range of observations which were each made for different purposes. Therefore, the selection function is complex and difficult to quantify. To control for this, the catalogue must be cross matched with existing surveys. The COSMOS survey is the ideal candidate for providing this. It is a well studied area of the sky, across a range of wavelengths. A primary band was the $F814W$ band in the COSMOS-\textit{HST} survey, which will have been previously searched for interacting galaxies in OR23.

Importantly, the COSMOS survey provides us with a large volume of ancillary data. From extensive photometry, to estimated galaxy parameters, to redshift estimates. The latter parameters are of the utmost importance to our study as the OR23 catalogue was a candidate catalogue. This is due to the difficulty in accurate predictions of galaxy interaction from morphology alone. Therefore, with the addition of photometric redshift information from COSMOS, likely interacting galaxies could be found. We describe the COSMOS survey further in the next section.

\subsection{The Classic COSMOS2020 Catalogue}
The COSMOS survey is a deep, wide field survey centred on coordinates ($+150.1192^{\circ}, +2.20583^{\circ}$). It covers a 2 square degree area about this central point and has been observed with many major observatories. These include space observatories such as \textit{HST}, Spitzer, GALEX, XMM, Chandra, Herschel and NuStar and ground based telescopes including Keck, Subaru, the Very Large Array and the European Southern Observatory Very Large Telescope. This gives it unprecedented wavelength coverage from the radio to the X-ray. The original survey overview is fully described in \citet{2007ApJS..172....1S}, which initially used \textit{HST}observations with the $F814W$ band \citep{2007ApJS..172..196K}, reaching a $5\sigma$ depth of 27.2(AB). Since this initial release, there have been many photometric catalogues from the COSMOS team using different fitting software across the photometric information they have available to them. This information has only grown as almost all major astrophysical observatories have observed this area.

One such catalogue is C20, containing a wealth of ancillary information for approximately 1.7 million sources. Each source has been analysed with well known astronomical software (\texttt{LePhare}, \citealt{1999MNRAS.310..540A, 2006A&A...457..841I}, and \texttt{EAzY},  \citealt{2008ApJ...686.1503B}). These provide estimates of the physical parameters of each source based on its measured broadband photometry. The parameters of interest in this work are the stellar masses and the SFRs of each source. For the stellar mass we use the best fit \texttt{LePhare} measurement and for the SFRs we use the best fit \texttt{EAzY} measurements. This breaks the degeneracy of the underlying spectral templates, but does introduce a need to remove sources with large differences in measured redshift of the two softwares.

Our sample of interacting galaxies contains only unique galaxies, but the C20 catalogue is not specifically a unique merger catalogue. We cross match between our sample and the C20 catalogues using a position search within $2^{\prime\prime}$ of our sample coordinates. Once we have identified the nearest C20 source for each interacting galaxy ID, we de-duplicate based on COSMOS2020 ID and redo the coordinate matching process with any duplicate matches. If no further C20 sources were within $2^{\prime\prime}$ of the source, then we classify the source as not in the C20 catalogue.

Once matched, we remove any sources with non-physical photometric measurements for the stellar mass or SFRs. We opt to institute a redshift cut of $z \leq 1.2$. Beyond this redshift, we find that identification of tidal features becomes difficult due to surface brightness dimming and we risk mis-identifying the stages of the interacting galaxies. Surface brightness dimming is a gradual process, which occurs with redshift by $\propto (1+z)^{-4}$. Therefore, the tidal features of an interacting system at $z=1.2$ will be $\approx$20$\times$ dimmer than a system at $z = 0.0$. We also use this cut as it matches the redshift cut applied to the environment catalogue we cross match with in Section \ref{data:environ}. Applying these cuts leaves a sample of 3,689 interacting galaxies - ${\sim}15$\% of the OR23 catalogue - within the COSMOS survey area.

\subsection{Secondary Identification}\label{sec:sec-ident}
\noindent The catalogue described in Section \ref{sec:or_cat} contains source coordinates and IDs of the a whole interacting system only. We, therefore, must manually identify the secondaries by applying the following three steps. First, a cutout of each source was created. These cutouts were from the COSMOS systems cutout service\footnote{Service: \href{https://irsa.ipac.caltech.edu/data/COSMOS/index_cutouts.html}{https://irsa.ipac.caltech.edu/data/COSMOS/index\_cutouts.html}}, selecting \textit{HST}-ACS tiles in the $F814W$ filter. Each cutout was 30$^{\prime\prime}$$\times$30$^{\prime\prime}$ (corresponding to 1001 $\times$ 1001 pixels). The original cutout from Section \ref{sec:or_cat} was also displayed next to the this larger cutout. Each source in the cutout was annotated with its COSMOS2020\_ID and photometric redshift. With this additional 3D information, we visually assessed interacting system giving one of four classifications: system disturbed but secondary could not be identified; secondary could be identified; cannot confirm galaxy is interacting; null redshift (photometric redshift was 0 or NaN); incorrect primary assigned. 

We defined the secondary galaxy(-ies) in the system as those with a measured redshift of $\pm$0.04 of the primary. Similar redshift cuts are often done when calculating environment parameters \citep[e.g][]{2006MNRAS.373..469B} or defining interacting galaxies by close pairs \citep[e.g][]{2022ApJ...940....4S}. A null redshift is defined as the primary galaxy out with our redshift limits, 0, or NaN. A minority of the cutouts we visually assessed were found to have the incorrect primary galaxy assigned, with the assigned primary being in the foreground or background compared to the interacting system. In these cases, we recorded the correct primary and secondary galaxy IDs and extracted the corrected information from the C20 catalogue.

\begin{figure*}
    \centering
    \includegraphics[width=0.95\textwidth]{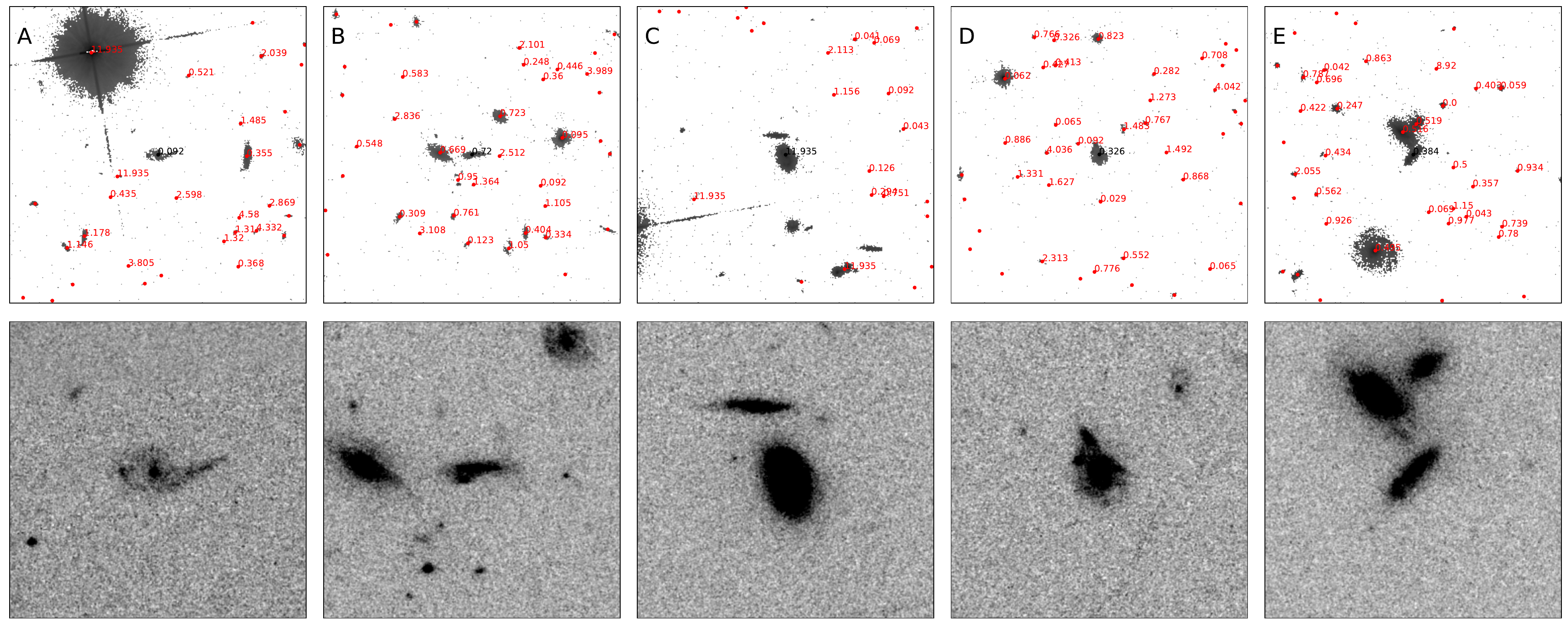}
    \caption[An example of each visual classification made on the cross matched sample.]{An example of each visual classification made on the cross matched sample. These images are 30" across using the COSMOS cutout service, selecting HST/ACS tiles as the basis for the observations in the $F814W$ filter. The photometric redshifts of each source are annotated in red. These are: (A) where the secondary could not be identified, (B) the primary had a clear secondary with a matching photometric redshift, (C) the primary could not be reliably classified as an interacting galaxy either due to a lack of photometric redshift information or morphological disturbance, (D) the redshift of the primary was null or (E) the incorrect primary was identified. Based on these classifications, we either add the secondary galaxies to the sample or we remove the contamination from it.}
    \label{fig:secondary_selection}
\end{figure*}

We find that of the 3,689 original systems cross-matched with COSMOS2020 2,283 could not have their secondary identified, 834 had a clear secondary, 446 could not be reliably classified as an interacting galaxy, 248 had a null redshift and 149 were the incorrect primary. Figure \ref{fig:secondary_selection} shows an example of each of our classifications.

\begin{figure}
    \centering
    \includegraphics[width=0.95\columnwidth]{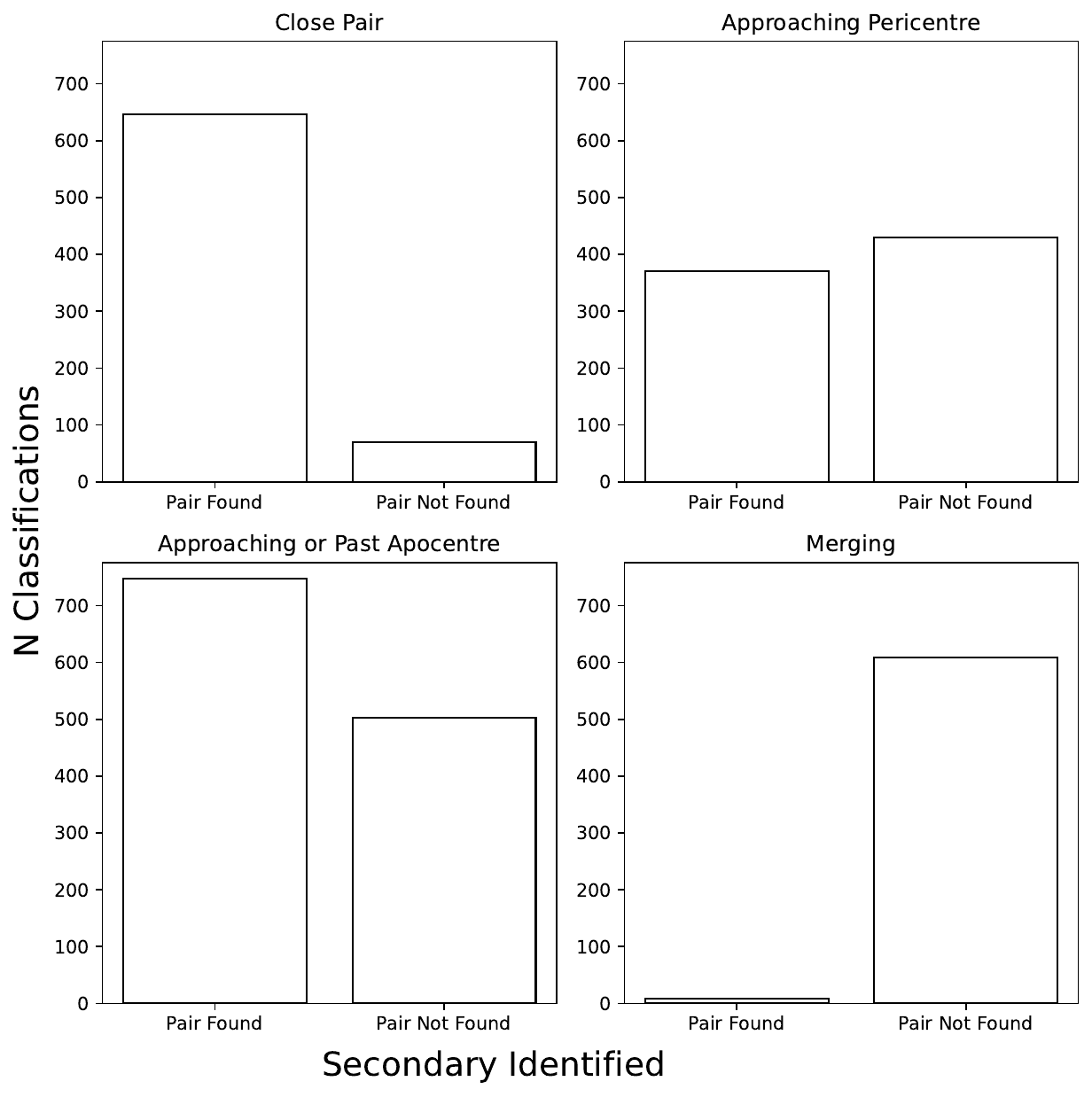}
    \caption[Where a secondary could be identified at different stages in the interaction.]{Number of interacting galaxies where a secondary could be identified at different stages in the interaction, often due to seeing. The reasoning for such disparity in secondaries identified is due to the relative distance each the secondary would be from the primary at each stage. For a close pair, we often found the secondary galaxy, but a minority of these were so close together that the entire system was given a single COSMOS ID. The same was true for those interacting systems approaching pericentre or merging. When the secondary was near apocentre, often it would be outside the cutout we were using for visual classification.}
    \label{fig:secondaries_found}
\end{figure}

While initially surprising that the majority of our systems could not have a secondary identified, we found that it was due to limitations in source detection or the how we had found the secondary galaxies. In the first instance, each potential secondary must have a COSMOS2020\_ID associated with it, which may not be the case if the two systems were not deblended correctly. The original source detection for the C20 was conducted using ground based observations. Therefore, if the sources were closer than the seeing of the observation - roughly 0.7", then they would not be resolved in the pipeline \citep{2016ApJS..224...24L}. However, in the \textit{HST}-ACS observations, we resolve them successfully. This was most pronounced when two systems were merging or at the pericentre of the interaction. This meant that the two systems would be classified under one COSMOS ID. Figure \ref{fig:secondary_selection} panel (C) shows an example of two systems being close enough together that they have been identified under a single COSMOS2020\_ID. Figure \ref{fig:secondaries_found} shows this disparity with the different types of interaction we observe.

Those galaxies which were found to be contamination (i.e., could not be reliably classified as interacting as they showed little tidal distortion or had no neighbouring systems at a matching redshift or having a redshift of 0) were removed from our sample. These systems were often individual galaxies with irregular morphologies of spiral arms or were very clumpy. There was also many systems that were at high redshift ($z > 1$) where the resolution of the cutouts meant that features could be difficult to discern visually. If the secondary galaxy could not be identified for the primary due to the above criteria, we did not include it in the paired sample.

\subsection{Finding Additional Systems}
\noindent As we use visual assessment of our cutouts with annotated 3D information, we are able to also confirm other interacting systems which were not present in the OR23 catalogue. Primarily, these extra galaxies are from systems which had more than two galaxies involved in the interaction, or were physically close pairs with little to no distortion. These galaxies would not have been found in the OR23 catalogue, as that prioritised morphological disturbance in its classification method. It also comprised other interacting galaxies that were added were low redshift systems which would have appeared to completely fill the cutout of the classification process. By looking at the larger COSMOS2020 cutouts, we are able to recover these galaxies and add them to this sample.

\subsection{Creating a Stellar Mass Limited Sample}
\begin{figure}
    \centering
    \includegraphics[width=0.95\columnwidth]{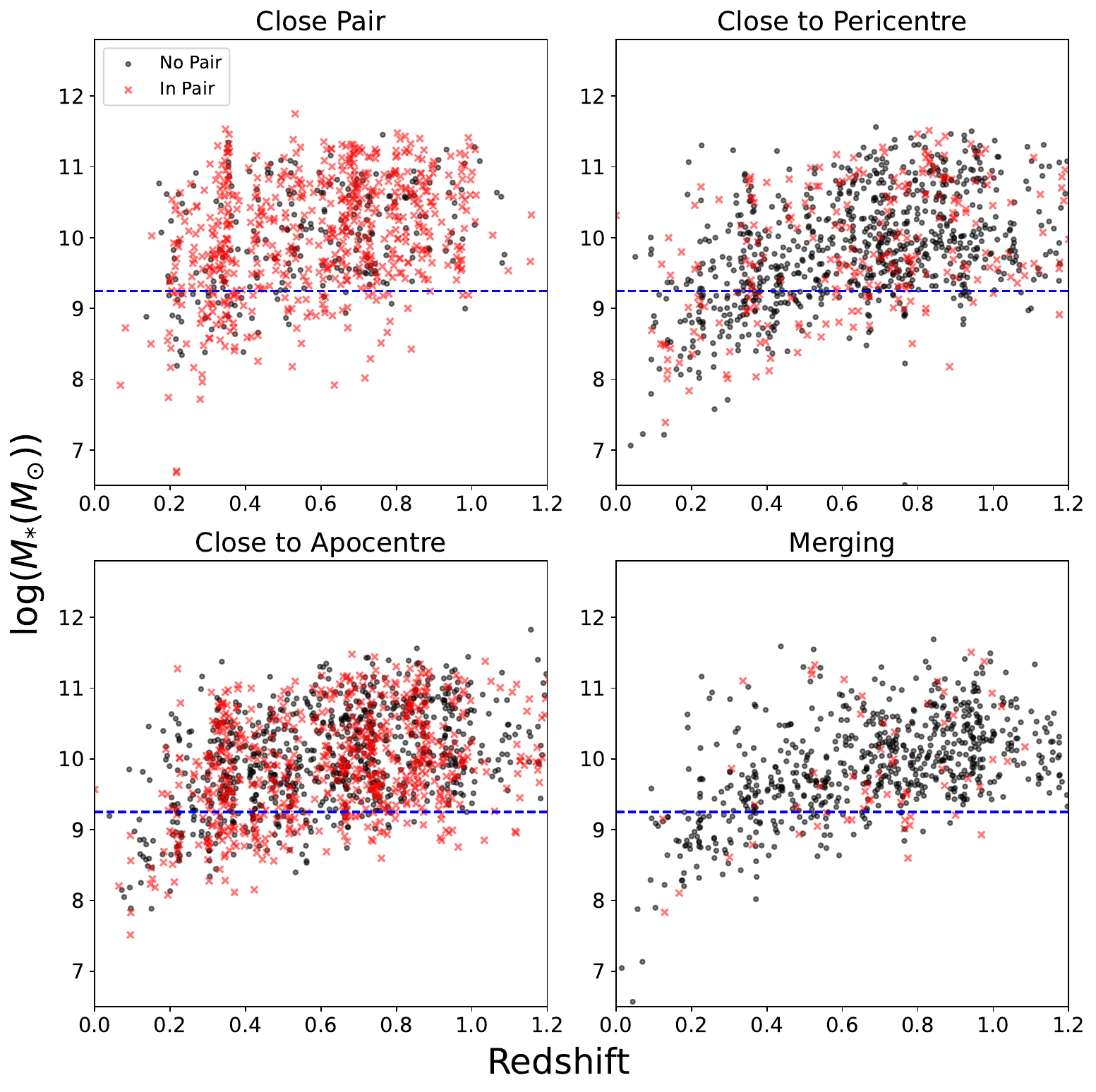}
    \caption[Redshift vs Stellar Mass distribution for each stage of interaction we have defined.]{Redshift vs stellar mass distribution for the four stages of interaction. We separate those systems with an identified secondary (in red) from those where no secondary could be found. The dashed blue line shows the stellar mass cuts we make for our stellar mass-limited sample and is set at $\log \left(M/M_\odot\right) \geq 9.25$. As shown here, the distribution of systems across redshift and stellar mass is consistent for all stages in our sample. This is important as we use tidal distortion and the existence as tidal features as a fundamental for our classification methodology. Therefore, we are likely not affected by this in our analysis. We find that our ability to identify pairs is primarily affected by stage and not redshift.}
    \label{fig:redshift_selection}
\end{figure}

\noindent We now investigate the distribution of our sample between stellar mass and redshift. Figure \ref{fig:redshift_selection} shows the resultant distribution from our sample. To ensure that our results are not biased by a redshift-dependent stellar mass or luminosity limit, we institute a cut in the stellar mass of the interacting systems. We elect to use a stellar mass cut of $\log\left(M/M_\odot\right) \geq 9.25$, shown by the blue dashed line in Figure \ref{fig:redshift_selection}. Note, if one galaxy in a pair is below this stellar mass cut, then we remove both galaxies from the sample. This ensures that our pair sample is also stellar mass limited. This is the lowest stellar mass system which is still observed in our the coalescing sample at $z = 1.2$. The second is that such a stellar mass cut is approximately the one made in the environment catalogue we describe later in Section \ref{data:environ}.

There is a limitation associated with making a stellar mass cut with redshift in this way. \texttt{LePhare} utilises SED fitting over broad band photometry to make estimates on the stellar mass. However, as the redshift increases and we begin to lose detection in different bands, the number of bands the SED can fit over reduces. This has the effect of increasing the error on our stellar mass estimate, and could skew or misrepresent it. To explore this, measure the change of the \texttt{LePhare} stellar mass probability distribution function (PDF) for different sources. The upper and lower percentiles of this PDF are recorded in the COSMOS catalogue, and we use this as a proxy for the change in shape of the distribution. We find that as we increase redshift, the mean and median sizes of the 68th percentile decreases. This means the PDFs are narrowing as we increase in redshift. Instead of this pointing towards an increase in accuracy, this is likely a descrease as the SED must now be fit over fewer bands. However, this change is minor, only changing from 0.5 - 0.3 over the entire redshift range. Therefore, we do not expect the COSMOS2020 stellar mass estimates to be majorly affected.

Further, we also identify systems where the stellar mass fit may be poor by accounting for the recorded error on the photometric redshift. To do this, we take the upper and lower 86th percentile measurements of the \texttt{LePhare} photometric redshift and measure the width of the PDF. We then divide this by the recorded median to calculate an estimate of the error of the measured photometric redshift on each source. We find for the majority of our sources, the error is less than 0.04. We remove any sources with an error larger that $\pm0.1$ error from this: ${\sim}200$ sources. Removing these sources mitigates the impact of close pairs by projection effects, rather than galaxies which are actually in physical proximity to each other. if we are highly contaminated by close pairs by projection, any signal from potential evolution with interaction stage would be weakened.

Throughout this work, we will use our stellar mass and redshift limited sample in our analysis. This acts as a volume limitation to our sample, and ensures we are not affected by evolution effects in our results. However, we have also conducted the same analysis presented in this work with the flux-limited sample and the qualitative conclusions of this work do not change. For a full summary of our selection process, and the source counts after each step, see Section \ref{sec:count-summary} and Table \ref{tab:source-count}. This Section also will summarise our environment and AGN selection.


\section{Methods} \label{method}
\subsection{Classifying Stage of Interaction}\label{sec:staging}
\noindent We split the dynamical time of interaction into four distinct stages. Each stage classification is based on the morphology of system. We have already encountered the four stages we will investigate in Figure \ref{fig:redshift_selection} and now provide a full definition here, and how we will refer to them for the rest of the work.

\begin{itemize}
    \item Separated: Systems which are well-separated with little to no morphological disturbance (Close pairs).
    \item Pericentre: Close systems showing morphological distortion while still in a pair or show a physical connection by tidal features.
    \item Apocentre: Well separated pairs with morphological disturbance or isolated galaxies with clear tidal features from unknown secondary.
    \item Merging: Highly disturbed systems with a secondary core present.
\end{itemize}

\noindent These definitions are similar to other recent works, such as \citet[][]{2022ApJ...937...97C} or \citet{, 2023ApJ...952..122G}. We have known since \citet{1972ApJ...178..623T} that the stage of the interaction in the dynamical time is linked to the tidal features formed. For instance, in long flybys of major interactions (apocentre in our case), long tidal tails are formed. In minor interactions, the secondary can become destroyed and lead to stellar streams forming about the primary galaxy. The latter are very difficult to detect, and require instruments capable of probing the low surface brightness universe. In these major interactions, tidal bridges can form between the two systems. The two systems must be close enough for these to form, and are therefore, likely at the pericentre of the interaction.

\begin{figure*}
\centering
\includegraphics[width=0.52 \textwidth]{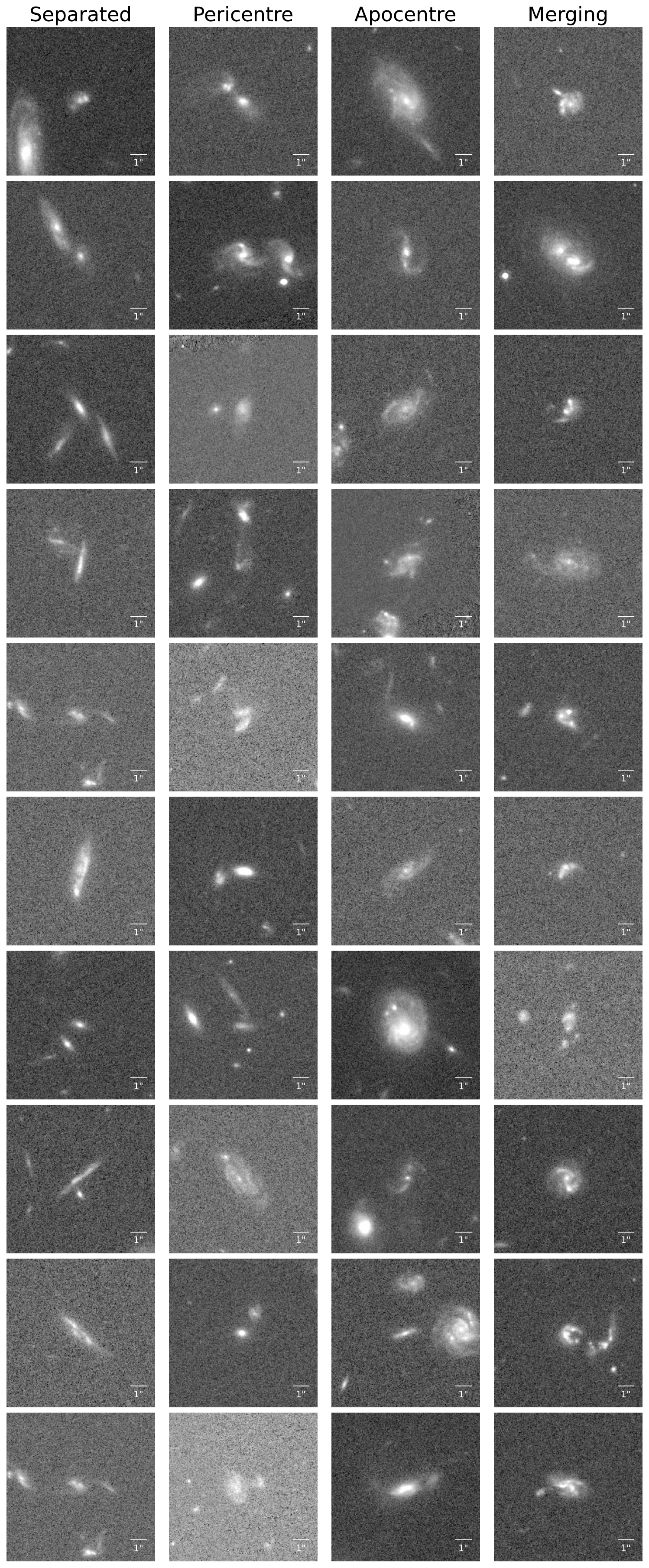}
\caption[Examples of the four stages of interaction galaxy sample into.]{Examples of the four stages interaction. Separated: A close pair with confirmed redshift matching. Pericentre: Two distinct systems interacting with tidal features forming. Apocentre: A tidally disturbed system with no secondary present, likely at apocentre. Merging: A galaxy with multiple cores while highly disturbed. At the final stage before coalescence.}
\label{fig:stages}
\end{figure*}

Figure \ref{fig:stages} shows ten examples of each stage. While we can estimate the stage from the morphology of the system, there are degeneracies associated with this approach. For context, we define a degeneracy as when the interacting galaxies may be at two or more parts of the dynamical timescale which cannot be separated without further information. {In this work, we only use the morphology and photometric redshift of the system to identify interacting galaxies. Additional information that would make our classifications much clearer would be the velocity information of the system. However, even with this information, this only provides insight into the line of sight velocity of the information. There is still opportunity for mis-classification of the stage of the system. To completely resolve this degeneracy, simulations would have to be employed. These could recreate the morphology of the system and 'back-propogate' the interaction to tell the time before or after the point of closest approach. While this is also limited by base assumptions, O'Ryan et al. (in prep) is building an algorithm to conduct investigate the feasibility of this on a set of major interacting galaxies \citep[for initial tests of this algorithm, see Chapter 4 of][]{d8bda841f36348319d7ddbb619148c91}. However, this is computationally expensive and difficult to apply to very large sets of interacting galaxies. Other examples of this approach also exist in the literature, for instance when looking at the merger history of individual systems \citep[][]{2025arXiv250211041B}. So, with these degeneracies in mind, we define the full breakdown of how we identify each stage.}

The separated stage captures the first approach of the two systems, where they are observed as a close pair. They have no morphological disturbance and, in our sample, are most often those galaxies with distinct disks. At this point, we would expect no change in the underlying processes of the galaxies from their control systems. The interaction has not taken place yet, and the two systems exhibit no morphological disturbance or tidal features. By definition, this stage requires an identified secondary galaxy and therefore has the highest number found. the degeneracy of this stage is not having velocity information of the galaxies, and therefore no way to identify if the galaxies are approaching or receding from another. It is possible for two galaxies to interact without exhibiting, or showing very subtle, distortion under certain interaction parameters.

The pericentre stage of the interaction is defined as the point where the two galaxies in the interaction are at or just passing the pericentre of the tidal encounter. At this point, we find the beginning of morphological disturbance, the initial formation of tidal features and tidal debris. This stage is primarily classified by the existence of tidal bridges, or closely overlapping tidal tails but also by highly disturbed overlapping disks. This stage is distinguished from the merger stage by the existence of a disk in both galaxies. Due to the two systems having to overlap or connect via tidal features, we find this stage contains the most systems with limited identification of the secondary galaxy. This stage is also highly degenerate in the context of the dynamical timescale of the interaction. Without further information, we are unable to define whether the galaxies involved at this stage are at the first, second, third, etc passage of the tidal encounter or if they are approaching or passed the point of pericentre.

The apocentre stage describes those interacting systems where the two disks are fully separated and distinct from one another. They have morphological disturbance associated with them, but do not require a secondary galaxy to be classified as this stage. These morphology disturbances most often take the form of disturbed disks with elongated tidal tails or streams about them (though, the former is in the minority). We distinguish these from the merger sample by requiring no secondary core to be present in the galaxy and for the galactic disk to be intact. We also use the existence of sheared material around the galaxy as evidence of an apocentre system, however remain cautious as this could be at the merging stage. Therefore, we use the existence of this debris and a nearby secondary within the necessary redshift bin to make the classification. However, the two galaxies in an interaction may have sufficient velocity to escape from each other and, therefore, their secondary could be beyond our COSMOS2020 cutouts. This is reflected in an even distribution of finding the primary and secondary in this stage. This stage also defines a large part of the dynamical timescale. It spans from the separation of the two galaxies after pericentre, to moving out to the apocentre of the interaction (or escaping with sufficient velocity), to falling back in towards the secondary galaxy again.

A further degeneracy exists in the orientation of the pericentre and apocentre systems. If the axis of the interaction is aligned with the line of sight - i.e. we are looking through one galaxy to the other - we are unable to identify whether the two systems are at pericentre or apocentre accurately. In such systems, it is very difficult to assume the redshifts would inform us of the line-of-sight separation. Classifying them in either category would therefore be difficult without more precise information. To calculate if this would affect our results, we create a toy model of interacting galaxies, this model is two points distributed in 3D. We create 10,000 of these points and then define an axis as the line-of-sight of the viewer. We then make an estimate in how many of these systems are aligned in the line of sight. Repeating this 10,000 times and measuring the standard deviation, we find this corresponds to  $(1.42 \pm 0.18)$\% of systems.

Finally, the merging stage represents the final step of a galaxy interaction. If the two galaxies do not have sufficient velocity to escape one another they will coalesce. We define this stage through extreme morphological disturbance of the system and the existence of a second core. The extreme morphology disturbance often takes the form of a completely destroyed disk, with of debris surrounding the galaxy. There can be a second highly disturbed system clearly merging into the disk, but not forming features like tidal bridges or tails. While we attempt to capture only pre- or ongoing-coalescing systems, it is important to note that this stage is degenerate to post-merger remnants which will also be accepted by our criterion and pericentre systems which are at the point of the closest approach. Post-merger remnants are systems where coalescence has been completed and with highly disturbed morphologies. The degeneracy with the pericentre point would be when the two galaxies are either actively passing through one another with the core of one galaxy within the other. They would also need to be hosting no extended tidal features, which occurs for a minority of interactions with specific interaction parameters. This is a relatively short time over the dynamical time, so will be a negligible fraction of systems.

\begin{figure*}
\centering
\includegraphics[width=0.95\textwidth]{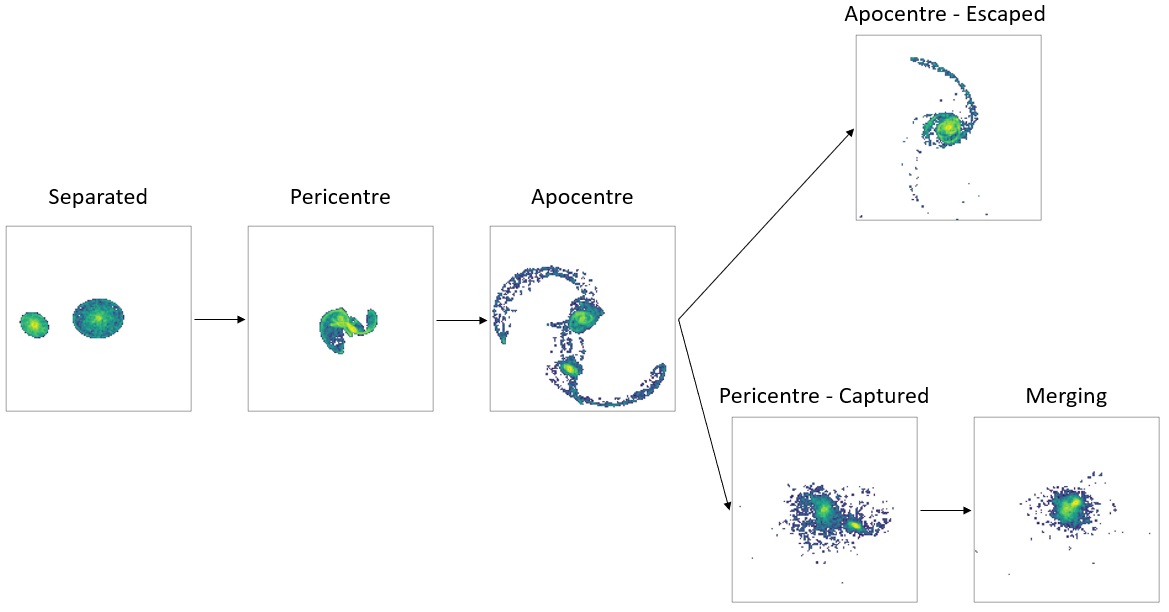}
\caption[The progression through an interaction using our stage definitions.]{The progression through an interaction using our stage definition. In the separated stage, two systems approach each other but exhibit no tidal features. This is before the point of closest approach has occurred. This is followed by the pericentre stage: the systems are approximately at their closest approach. Clear tidal features exist with major disturbance in the two disks. This is followed by the apocentre stage, where two distinct cores with clear tidal features. However, after this point, there are two outcomes to the system depending on the velocities. If the secondary has the escape velocity, the system will remain an apocentre stage interaction until the tidal features dissipate (and no longer are in our sample). If they do not, the system will return to the pericentre stage of the encounter and then begin to coalesce in the merging stage. Images are from the Advanced Python Stellar Particle Animation Module interacting galaxy algorithm described in O'Ryan et al., in prep and based on the stellar particle animation module algorithm described in \citet{2016A&C....16...26W}.}
\label{fig:illustration}
\end{figure*}

We can now we can put our classification system into the context of the entire dynamical timescale. We would expect the galaxy pair to begin in the separated stage and progress to the pericentre stage in the early times of the dynamical timescale. Then, dependent on the velocity of each galaxy, the pair will either progress into the merging stage and begin to coalesce or the apocentre stage. This change from pericentre to apocentre can then take two further branching paths. If the galaxies have sufficient velocity, the apocentre stage will be their end state until the tidal features slowly dissipate. If they do not have sufficient velocity to escape, the system will move from the apocentre stage back into the pericentre stage. This could happen for many cycles until the galaxies enter the merging stage and coalesce. Figure \ref{fig:illustration} shows the branching paths that the galaxies can take through each stage. These images are created using the Advanced Python Stellar Particle Animation Module restricted numerical simulation, described in O'Ryan et al., in prep, and are for illustrative purposes only here.

Rather than classifying the stage of an interaction based on morphology the projected separation of the two systems is used. To explore the difference between them we measure the projected separations of our confirmed galaxy pairs. We measure this by taking the average of the best fit photometric redshift of the two galaxies, and converting their angular separation to a physical one. The most distinct projected separation ($s_{\mathrm{proj}}$) in stages is between the pericentre and apocentre stages. Here, we see that the pericentre stage is dominated by systems with $s_{\mathrm{proj}}<$35kpc while the apocentre stage is dominated $25 \leq s_{\mathrm{proj}} \leq 100$ kpc. 

Those galaxy pairs in the pericentre stage with large projected separations are pairs which are very large in angular size, while still overlapping or morphologically linked. The evolution of the tidal features in these examples also plays a role in this overlap. Different mass ratios, gravitational potentials and orbital parameters will lead to different distances in each system for defining our change from a pericentre and apocentre system. However, in terms of the dynamical time, these points are the same and we would expect the effects on the galaxies to be similar. In a way, we are normalising over the dynamical time by only considering the morphology and evolution of the tidal features over the projected seperation only.

\begin{figure}
    \centering
    \includegraphics[width=0.95\columnwidth]{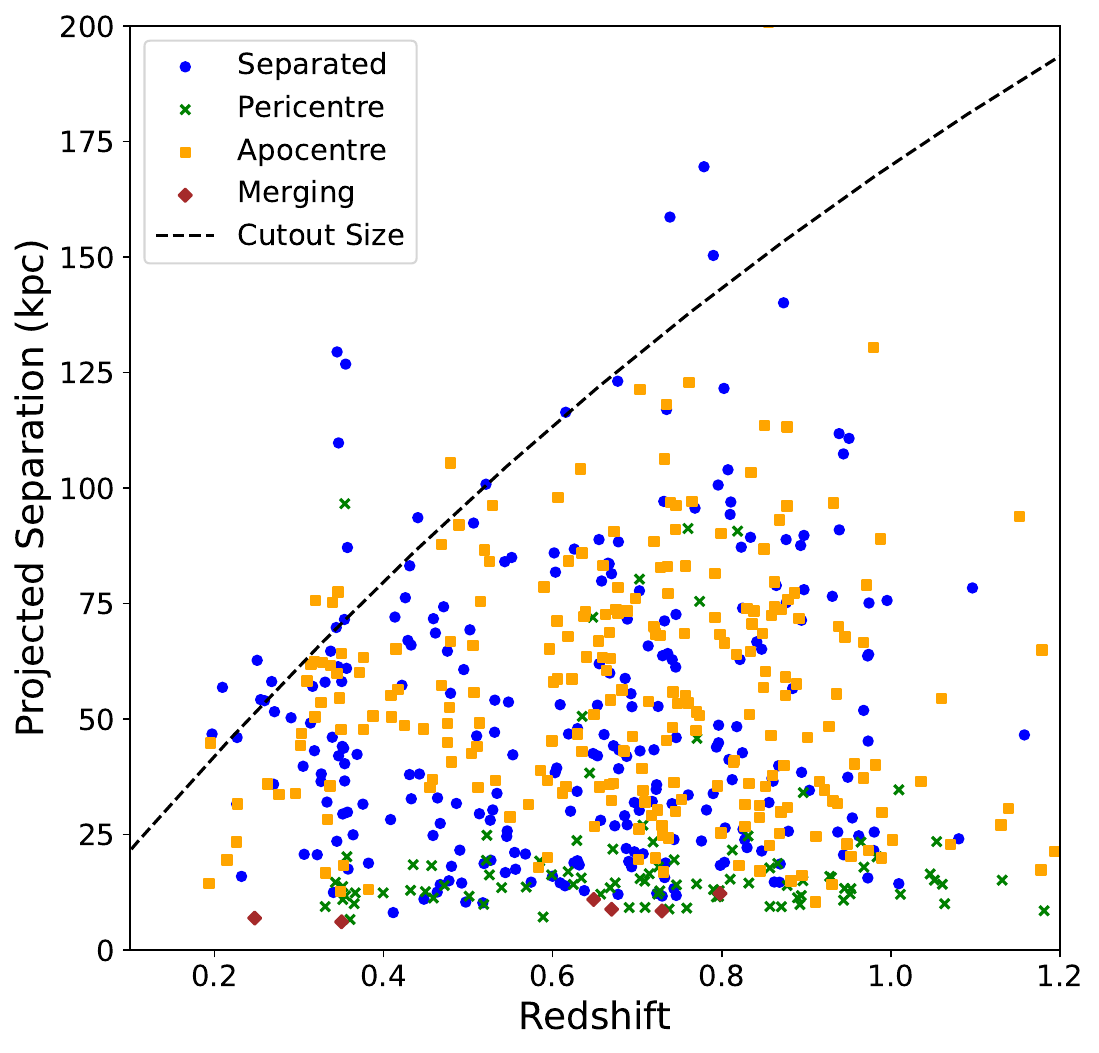}
    \caption[The measured distribution of projected separation with redshift.]{The measured distribution of projected separation with redshift. The dashed line shows the projected size of the cutouts used for classification of interacting galaxies in OR23. We have made redshift cuts out to $z = 1.2$ and we investigate the limitations on the projected separations we can successfully identify across our volume. We find that we can successfully identify secondary galaxies and their projected separations down to 10kpc to $z=1.2$. This is true across defined interaction stage. We find, because of the defined size of our cutouts, the limitation of finding the secondary galaxy is at low redshift.}
    \label{fig:proj-seps-limits}
\end{figure}

Figure \ref{fig:proj-seps-limits} shows the overlap between our four stages in projected separation. We find we identify fewer pairs at low redshift where a larger projected separation represents a larger angular separation on the sky. This is a limitation of using visual confirmation of the secondary within a cutout of limited angular size. This is marked by the black dashed line as the maximum size of the original OR23 cutout with redshift. We do find some systems above this limit, but this is due to our identification methods in this work. Many of these systems were individually identified as interacting galaxies in the OR23 catalogue and, using the larger COSMOS cutout of this work, we were able to pair them together into their larger system. A subset of these ($\approx5$) were newly identified systems which were at the same redshift as the OR23 identified galaxy. These systems exhibit little tidal disturbance, and hence were missed in the original OR23 catalogue. We also find at low redshift ($z < 0.2$) we only identify pairs of galaxies with projected separation below $50$kpc. Towards our limiting redshift, however, we are able to identify galaxy pairs down to a projected separation of 5kpc, as well as out to 200kpc. 


\subsection{Matching to Environment Catalogue}\label{data:environ}
\noindent It is well understood that galactic environment has a direct impact on the observed SFR of a galaxy. A galaxy in a cluster environment has, on average, a higher SFR than those in the field \citep{2006MNRAS.373..469B}. Thus, if any of our stage classifications are biased towards one environment or another, it could impact our results.

There is no measure of the environmental density in the C20 catalogue. Techniques for measuring this vary between works, but examples include the Nth-nearest neighbour \citep{2006MNRAS.373..469B}, different Bayesian metrics \citep{2008ApJ...674L..13C} or estimating it from Voronoi Tesselation \citep{2021inas.book...57V}. However, in this work, we use the existing environmental density catalogue produced by \citet{2017ApJ...837...16D}. This catalogue was created specifically from the COSMOS2015 catalogue, and has a measured density for all sources with stellar mass $\log\left(M/M_\odot\right) \geq 9.6$ and $z \leq 1.2$. These limits were motivated by conserving completeness of their sample for accurate enviroment measures. \citet{2017ApJ...837...16D} calculate not only the density, but also the density parameter $\delta$ and assign each source to a field, filament or cluster classification. For a full description of how they calculate the environment and density field see \citet{2015ApJ...805..121D, 2017ApJ...837...16D}, we will only briefly describe it here.

To build the density field throughout the COSMOS field, \citet{2017ApJ...837...16D} first construct a set of overlapping redshift slices. Within each slice, a subset of the galaxies are selected such that the median of the probability distribution function (PDF) of their photometric redshift is within it. Then, from this subset, they calculate the weighted surface density within the redshift slice. The weighting is based upon the PDF of the photometric redshift present within the redshift slice. These weights significantly reduce projection effects. They then apply a weighted adaptive kernel smoothing using a 2D Gaussian kernel whose width changes based on the found local density of galaxies. Once this density field is created, the density around the sample galaxies can be interpolated across the density field based on the angular position.

The result of this process, and the cuts defined previously, is a catalogue of environment densities for $\approx$45,000 galaxies. We remove any sources which are flagged as uncertain from the catalogue providing us with $\approx$39,000 sources with which to cross match our volume limited sample. We apply the same stellar mass criteria as \citet{2017ApJ...837...16D}, only considering those systems with a stellar mass $\log\left(M/M_\odot\right) \geq 9.6$. To cross match with our sample, we use the COSMOS2015\_ID which exists in both the C20 catalogue and the \citet{2017ApJ...837...16D} catalogue. Upon applying the stellar mass cut to our sample, we find 2,800 matches to the \citet{2017ApJ...837...16D} catalogue.

\subsection{Classifying AGN}\label{sec:agn-clsf}
\noindent We will also investigate the effect of interaction stage on AGN activity throughout our sample. As the C20 catalogue does not contain the relevant classifications, we turn to the Chandra COSMOS Legacy Survey Multiwavelength Catalogue \citep{2016ApJ...817...34M} and the COSMOS VLA 3GHz survey \citep{2017A&A...602A...6S, 2017A&A...602A...3D}. Both of these catalogues span the COSMOS survey area. The Chandra survey spans the X-ray range of wavelengths while the VLA 3GHz survey is used to find the radio AGN through our sample. 

Our matching process is similar to what we have previously described. We use the identified COSMOS2020 source coordinates and select the nearest source with in a 2" matching radius. We first find radio AGN using the VLA 3GHz catalogue. At every step, if we find a match in the relevant catalogue, we remove it from subsequent searches in other catalogues and take the first classification as the correct one. A flag is present in C20 catalogue for a matching Chandra ID. We use this to match between our sample and the Chandra survey directly. We use the AGN and star forming galaxy (SFG) classifications present in the VLA catalogue. For the Chandra catalogue, we define an AGN as having an X-Ray luminosity greater than $10^{42} \text{ergs s}^{-1}$. This luminosity cutoff follows the definition of \citet{2017MNRAS.465.3390A}.

Applying our matching criteria, we find 1,059 matches in the VLA 3GHz survey and 57 in the Chandra survey. From existing flags within the catalogues, these were split into 812 star forming galaxies and 304 AGN. We also investigate cross matching with the MPA-JHU catalogue \citep{2003MNRAS.341...33K, 2004MNRAS.351.1151B, 2007ApJS..173..267S}, all matches found were already represented by the VLA and Chandra surveys. We also use the COSMOS XMM-survey and, again, find no new sources to add to our sample.

\subsection{Defining a Control Sample}
We also define a control galaxy sample with which to compare our interacting galaxy sample to. For each galaxy in our stellar mass and volume limited sample, we find a stellar mass- and redshift-matched control galaxy. We draw these galaxies from the C20 catalogue. All galaxies within $\pm0.01$ dex of our samples' stellar mass are selected from the catalogue, and within a $\pm0.01$ redshift slice. We apply these criteria across the whole C20 catalogue, often giving us a list of potential control galaxies we could use. 

To reduce this list, we then apply an environmental criterion to our control galaxies. If an environment measure is available, then we take the control galaxy with the in a matching environment classification. If all potential control galaxies in the list have an environment classification, but none match the interacting galaxy then we do not use any of them. If neither the interacting galaxy or control galaxies have environment classifications (due to our samples lower mass cutoff than the \citet{2017ApJ...837...16D}) then we keep each potential control galaxy in the list. If more than one potential control galaxy matches the environment classification of the interacting galaxy, we apply a final angular separation criterion.

Using these criteria, the majority (${\sim}70\%$) only have a single control galaxy that fits the stellar mass, environment and redshift matching criteria we have applied. For the remaining ${\sim}30\%$, we apply an angular separation criteria to them. We select the control galaxy in this list as the closest in angular separation from the interacting galaxy. This increase the likelihood that the two systems will be in the same environment - if the interacting galaxy is one without an environment classification - but also allows us to recover the same control galaxy without any further information. We will be comparing our interacting and control samples' relative SFR distribution and AGN fraction, and therefore, we do not consider either of these parameters in our matching process. From this, we find control galaxies all but one of the galaxies in our interacting sample.

\subsection{Summary of Selection Process and Source Counts}\label{sec:count-summary}
We summarise each of the steps we have described here in a single table. Table \ref{tab:source-count} shows the breakdown of the number of sources added or removed at each selection step, and we introduce the numbers here. First, we cross match 3,689 sources from the OR23 catalogue to C20. We then conduct visual inspection of each source with photometric redshift measurements annotated. Using these allow us to confirm secondary galaxies, or other interacting systems, in the COSMOS field, finding 803 more systems. This gives us a total of 4,181 interacting galaxies, of which 1,690 were confirmed to be in a pair (or more). We then cross match our 4,181 sources with the \citet{2017ApJ...837...16D} environment catalogue of the COSMOS2015 survey and find 2,225 of our sources have environment measurements. We then apply a volume limit to the sample, applying stellar mass cuts of $9.25 \leq \textrm{log}_{10} M_{*}/M_{\odot} \leq 12.5$ and a redshift cut of $0 < z < 1.2$. This volume limited sample contains 3,162 sources of which 1,132 are in a pair and 2,214 have associated environment classifications.

This volume limited sample was then cross matched with the VLA and Chandra catalogues which existed in the COSMOS archives and the MPA-JHU catalogue. Of our flux limited sample of 4,181 sources 1,106 sources were cross matched with these catalogues. Applying the volume limitation gives 802 sources with classifications of AGN or SFG.

\begin{table*}
    \centering
    \resizebox{1.00\textwidth}{!}{\begin{tabular}{|c|c|c|c|c|}
        \hline
        Selection Process & Sources  & Paired Galaxies & Environment Measurements & AGN \& SFG Classifications \\
        \hline
        In COSMOS & 3,689 & - & - & - \\
        Visual \& Secondary Identification & 3,829 & - & - & - \\
        Deduplicating Sample & 3,547 & - & - & - \\
        Extra Found Systems & 4,181 & 1,690  & 2,225  & 1,116 \\
        Volume Limited & 3,162 & 1,112 & 2,214 & 802 \\
        \hline
    \end{tabular}}
    \caption[Source counts at each stage of our selection process.]{Source counts at each stage of our selection process.}
    \label{tab:source-count}
\end{table*}

\subsection{Visual Classification: Sources of Contamination}
\noindent Throughout the description of our sample, we have identified interacting systems by a combination of visual classification and the best-fit photometric measurements from the C20 catalogue. This introduces limitations which could bias or contaminate our sample. We identify three serious areas of bias or contamination: 1) error inherent from using only photometric redshift measurements, 2) failure of identification of tidal features at higher redshifts due to surface brightness dimming and reduction in angular size, and 3) mis-identification of tidal features as disturbances caused by other processes. In this subsection, we will address each of these points and quantify how they affect our sample. We will also discuss how we mitigate these effects where appropriate.

First, there is always an inherent error in using the photometric redshift measurements of galaxies. This was briefly mentioned when describing the pericentre stage systems. There, we found ${\sim}50$ systems which were clearly morphologically disturbed with linking tidal features, but their photometric redshifts were such that they couldn't be interacting. It is also possible that we have missed pair identifications in the separated stage (where no tidal features are expected) due to incorrect photometric redshifts. We wish to quantify both of these effects. \citet{2022ApJS..258...11W} quantify the error in their photometric redshift calculations (which we use here) as $<1$\% for our redshift and flux range. They quantify the scatter in the distribution of photometric to spectroscopic redshifts as $0.025(1+z)$ within our redshift and flux range. This leads to a maximum error on our photometric redshifts in this range of $\pm0.06$. 

To investigate the affected systems, we therefore look at the redshift distribution of our paired sample. During the visual classification, we instituted a cut that each interacting pair must be within a photometric redshift difference of $\pm0.04$. To estimate our completeness, we apply a bootstrapping test to our sample and see the number of systems we may have removed from our sample with this cut. We create a simulated paired sample of 10,000 systems and apply an error measurement based on the photometric redshift errors. This gives a paired sample across our redshift sample at slightly different redshifts. We then apply our cut, and check how many systems we would remove due to the photometric error from the C20 catalogue. We find that we would retain $\approx91$\% of the sample with this cut, given the photometric redshift error. This is a lower limit on what we have retained, as we have not considered possible tidal features that would have caused us to keep the system in our sample. Therefore, we are unlikely to have removed a large percentage of interacting galaxies due to the photometric redshift error and the cuts we have applied.

The second limitation of our approach is the failure of identifying interacting systems at high redshift due to missing the tidal features of the system. This would be a combined result of the affects of surface brightness dimming of their disks and tidal features at high redshift and while also having a small angular size. The former issue is mitigated by our limiting of our sample to be within a range of $0 \leq z \leq 1.2$ and with our stellar mass limit of $9.25 \leq \log\left(M/M_\odot\right) \leq 12.5$. The latter is more difficult as we have a standardised cutout size which does not depend on the redshift, or angular size of the system. This means that if a system were at redshift 1.2, our pixel resolution corresponds to roughly $0.5\rm{kpc~pix}^{-1}$ limiting the tidal features we would be able to identify. With such a scale we are able to identify extended tidal features away from the galactic disk such as tidal arms, but lack the resolution to identify those features which would be much closer to the disk. For example, those interacting systems which have formed features such as shells, small tidal bridges or stellar streams. This results in, at high redshift, retaining sensitivity to the separated, apocentre and merging stage interacting systems while lacking the required resolution to identify more intermediate tidal features at the pericentre stage.

We look at the change in found fraction of micro, minor and major interactions across our redshift range. The effect of surface brightness dimming will be related to the type of interaction we are observing, and the resultant flux distribution of the tidal features. We split our pair sample into its constituent interaction types based on the mass ratio. We define the mass ratio such that the primary galaxy contains the highest stellar mass in the pair. If there is more than two galaxies involved in the interaction, we take the primary galaxy to be the galaxy with the highest stellar mass in the system. We define three different interaction types based on the mass ratio following \citet{2008MNRAS.391.1137L}: micro (where the mass ratio is less than 1:10), minor (where the mass ratio is between 1:10 and 1:3) and major (where the mass ratio is greater than 1:3). We would expect that major interactions would not be changed by surface brightness dimming across redshift as these interactions create strong tidal features. The effects of micro interactions, such as creating stellar streams about the primary galaxy, will be lost.

\begin{figure}
\centering
\includegraphics[width=\columnwidth]{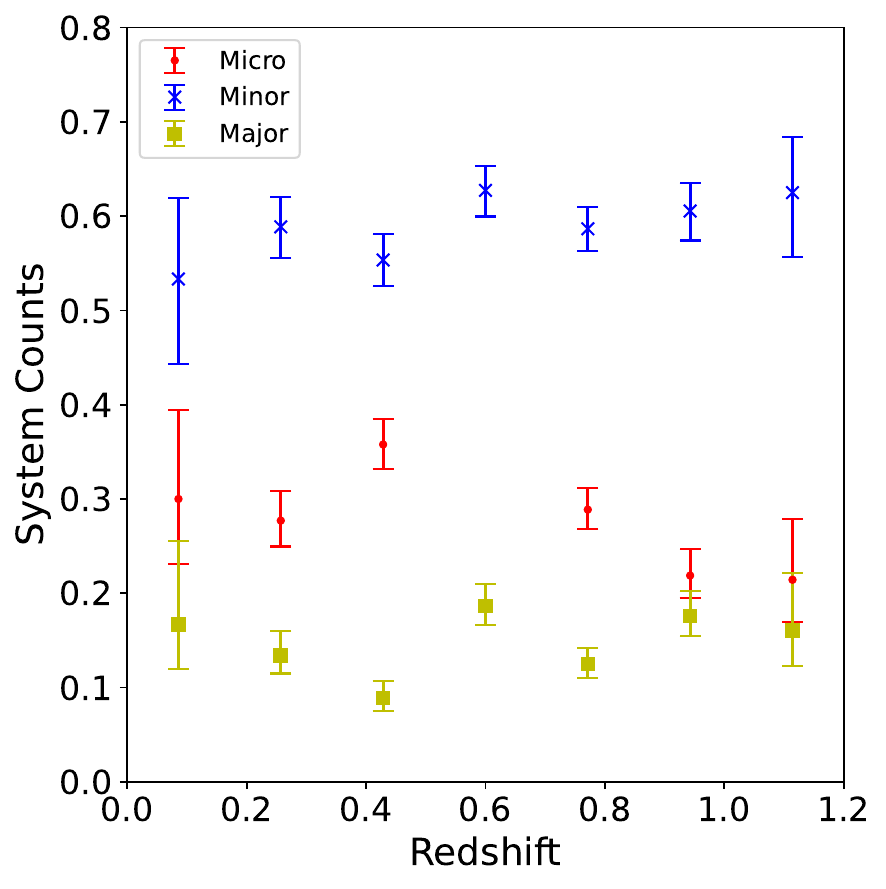}
\caption[The change in found interaction type across our redshift distribution.]{The change in found interaction type across our redshift distribution. We define interaction types from the mass ratio of paired galaxies in our paired sample. The change in the fraction found of different interaction types with redshift. As expected, the minor and major interaction fractions remain consistent with redshift but the micro interaction fraction declines. This is a result of surface brightness dimming on the tidal features formed in the interactions. In a micro interaction, we would see the complete destruction of the secondary into stellar streams about the primary galaxy. Such a tidal feature would be in the low surface brightness regime, and quickly be lost at increasing redshift. However, the change in fraction we find across our redshift range is not high enough to alter our conclusions.}
\label{fig:mass-ratio-limitation}
\end{figure}

Figure \ref{fig:mass-ratio-limitation} shows the change in the fraction of each interaction type we find in each redshift bin. For the minor and major interactions, we see that there is little change in the fractions across our redshift distribution. However, as expected, with micro interactions we see a gradual decline in the found fraction with increasing redshift. This shows the loss in sensitivity we have to low surface brightness features due to the previously mentioned surface brightness dimming. At this maximum redshift, tidal features would be $\approx$20$\times$ dimmer than at $z=0.0$. While we do begin to see the effects here, it does not affect our major or minor interacting sample as these remain reasonably consistent with redshift. Therefore, we do not attempt to increase or decrease the redshift cut.

The final limitation of our approach is our reliance on visual classification to identify interacting galaxies. Besides the issue of close pairs we have made the assumption that the morphological disturbance of the systems is due to interaction, and not from other processes. The most obvious process which could mimic tidal disturbance is that of ram pressure stripping (RPS). RPS is an effect of the galactic environment stripping out the gas within a galaxy, and causing the formation of debris about the galactic disk. It can also cause major disturbance and irregular morphology, which could be misclassified as a galaxy in the apocentre or merging stages of our sample. The environment where RPS is most prevalent is a cluster environment. Therefore, if we are highly contaminated to identifying RPS galaxies as interacting galaxies, we would see a bias in the environment of our sample towards galaxy clusters. We use the sample matched to \citet{2017ApJ...837...16D} as described in Section \ref{data:environ}.

It is important to note that \citet{2017ApJ...837...16D} has a higher stellar mass cutoff than we have implemented in our underlying sample. Therefore, we show the density of all systems $\log\left(M/M_\odot\right) \geq$ 9.6. The environment which would have the most effect upon the SFRs of our sample is a cluster environment. However, the fraction of our sample within a cluster is never greater than ${\sim}20$\%. Therefore, the lack of information here does not have a large impact on this measurement. 

\begin{figure}
    \centering
    \includegraphics[width=\columnwidth]{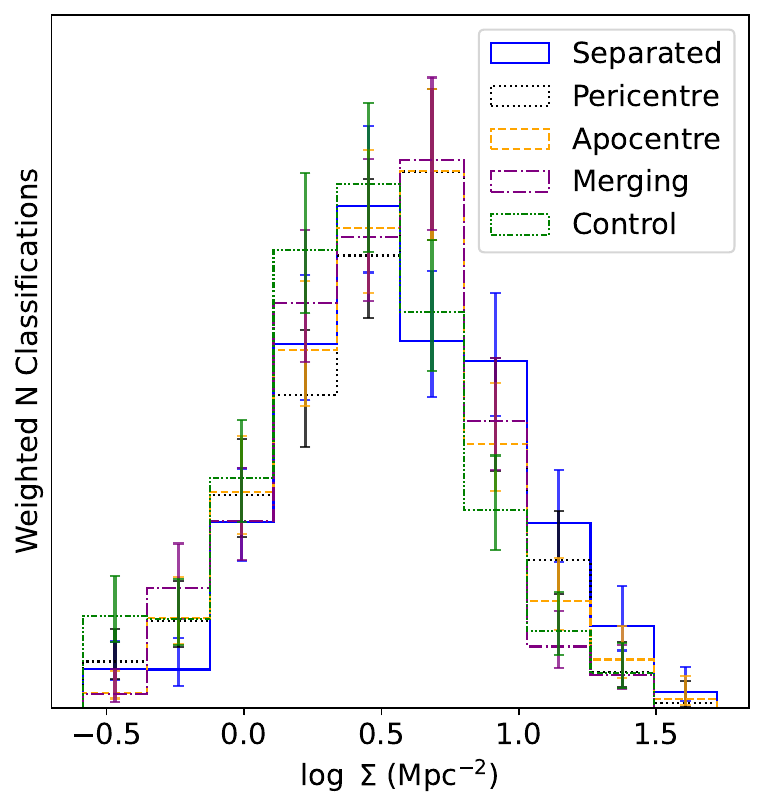}
    \caption[The distribution of environment classifications through our sample.]{The distribution of surface density that each of our sources resides within weighted by stellar mass. This is split into different classifications of interaction. While there is some variation, there is no major difference between the environmental distribution of each of classified interactions. We find that our separated stage resides in slightly higher environments, but only to $<1.5\sigma$. Therefore, we can exclude environment as a source of differing star formation distributions between each interaction classification.}
    \label{fig:dens-sfr-mass}
\end{figure}

To investigate this, Figure \ref{fig:dens-sfr-mass} shows the environment classification in our reduced stellar mass-SFR distribution. We conduct a weighted Kolmogorov-Smirnov \citep[KS-test;][]{an1933sulla} test to compare the distributions. The KS-test is excellent at comparing different weighted distributions and indicating if they are drawn from the same parent sample. These tests confirm they are consistent with being drawn from the same parent sample, with the resultant $p$-values $\approx1$ the distributions.

We have identified the merging stage as those systems with a disturbed disk and containing more than one core. Another such system that could be recognised by this morphological description is that of a clumpy galaxy. Clumpy galaxies are systems with internal regions undergoing intense star formation. These regions appear like multiple `cores' throughout the galactic disk which are actually clumps of young stellar populations. To avoid such contamination, we apply multiple criteria to mitigate the potential impact they might have. First, our redshift range spans a sufficient range that the found fraction of clumpy galaxies declines rapidly compared to the merger rate \citep{2022ApJ...931...16A}. This is particularly true at $0.15 \leq z \leq 1$. Previous works have found the peak of the fraction of clumpy galaxies at $z\approx2$ \citep{2014ApJ...786...15M, 2018ApJ...853..108G}, beyond our redshift range. There are also two morphological distinctions that we use to our advantage to discern between merging and clumpy galaxies. Clumpy galaxies often exhibit more than one clump at different radii across the galactic disk \citep[with][finding a mean of 3.16 clumps per galaxy]{2022ApJ...931...16A}, making them easy to differentiate from a second core which would be close to the centre of a perturbed disk. We therefore remove potential merging galaxies which appear to have more than one `core' at different places of a non-perturbed disk (${\sim}25$).

However, the tradeoff for better identification is the existence of two cores in the galaxy can also cause some inaccuracies and biases with SED fitting, an effect of failed deblending \citep{2009ApJ...696..348W, 2024MNRAS.530L...7H}. While it found that SED fitting algorithms find the correct SFR, often the stellar mass of the system is under-estimated. To mitigate potential failed deblending, we investigate the merging sample with the Farmer catalogue from COSMOS \citep{2022ApJS..258...11W}. This catalogue has a novel deblending scheme which differs from C20 - the \texttt{Tractor}. We use this catalogue to examine if we have major changes to our distributions between deblending algorithms. We find the merging stage distribution to be the same between the two catalogues. Therefore, we do not expect our results to be a result of failed deblending in the C20 catalogue.

\section{Star Formation and AGN Evolution with Interaction Stage}\label{results:SF_stage}
\subsection{Controlling for Interaction Stage}\label{results:int_stage}
\noindent We show the results of breaking down our sample into stages with relation to the SFR and stellar mass using the measurements in the C20 catalogue. We then use our subsample of galaxy pairs to recover the relationship between projected separation and star formation enhancement (SFE). We then further break this measure down into its component stages.

\begin{figure}
    \centering
    \includegraphics[width = \columnwidth]{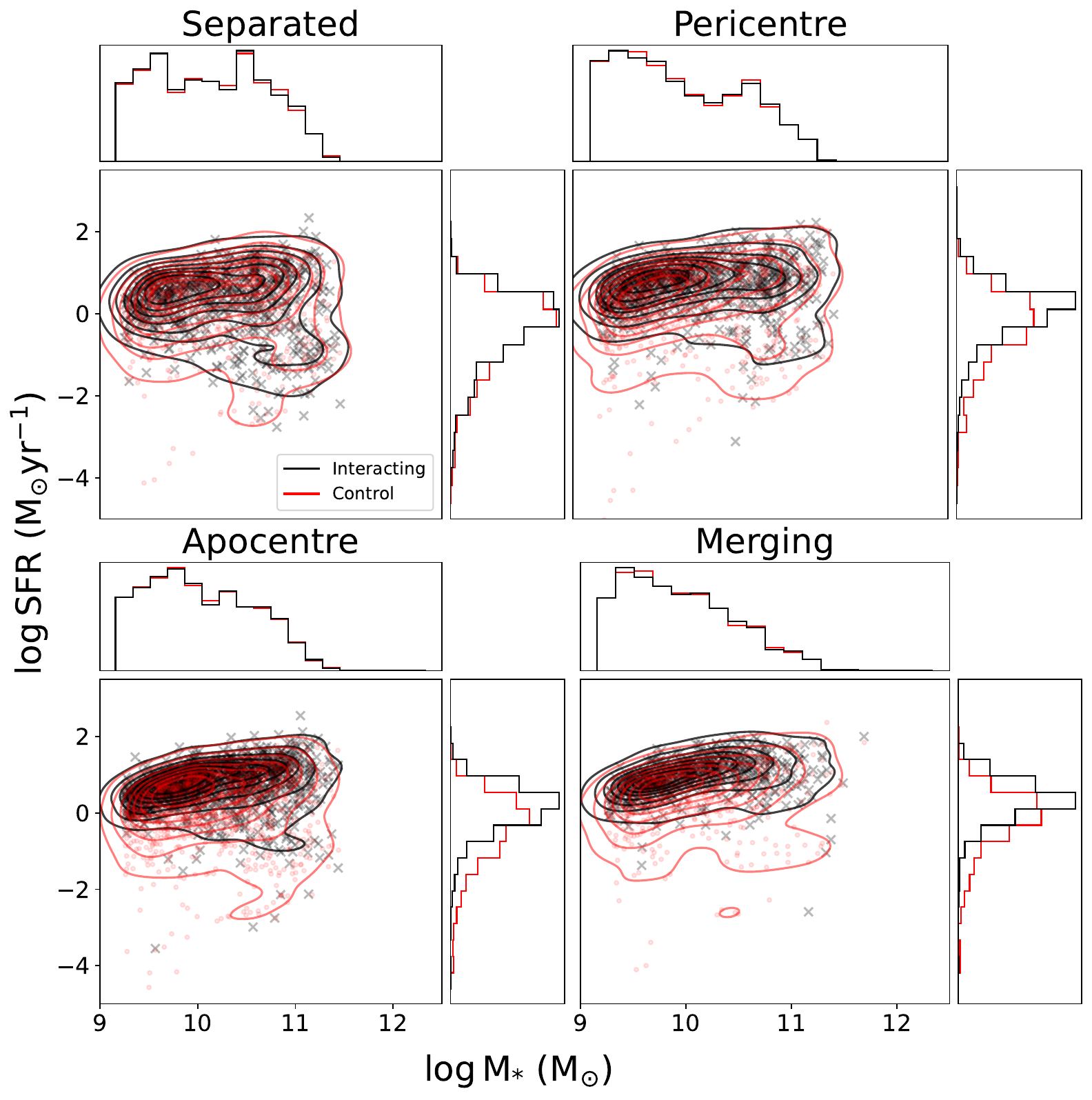}
    \caption[The \texttt{LePhare} stellar mass against the \texttt{EAzY} SFR across the different stages of the interaction.]{The \texttt{LePhare} stellar mass against the \texttt{EAzY} SFR across the different stages of the interaction. The black contours are 8 levels of density of the interacting sample in each frame. The red contours are 8 levels of density of our control sample in each frame. \textit{Top left}: The stellar mass and SFR of the separated stage of our sample. Here, the interacting galaxies are close pairs with little to no morphological disturbance. There are clearly two populations here: a main, star forming sequence forming the main population and a smaller red sequence. This is true of both populations. \textit{Top right}: pericentre stage of the interaction, where the two interactors are close to pericentre. The star forming sequence remains, but the red sequence is reduced significantly, while remaining consistent for the control sample. \textit{Bottom left}: apocentre stage of the interaction, where the interactors are close to apocentre or escaped. Here, we see the almost complete disappearance of the red sequence in the interacting population, while it remains for the control sample. \textit{Bottom Right}: merging stage of the interaction, where the two systems are close to or have coalescence. The red sequence of galaxies has completely disappeared in the merging sample.}
    \label{fig:sfr-mass}
\end{figure}

Figure \ref{fig:sfr-mass} shows the breakdown of stellar mass and SFR with stage. The black contours and markers show the change in distribution of our interacting sample while red shows our control. We find clear evolution in the SFR from the separated stage through to the merging stage when compared to the control sample. In the separated stage, where the galaxies are distinct from one another with no clear morphological disturbance, we clearly see two populations of galaxies: the blue, star-forming galaxies and the red sequence of quenched galaxies. The blue contours in Figure \ref{fig:sfr-mass} show increasing number density in the population into the blue cloud. In the pericentre stage, when the galaxies are actively interacting and overlapping, this red sequence remains but is highly diminished while there is no change in the blue cloud. The apocentre stage shows a similar effect, where the red sequence reduces again before disappearing in the merging stage.

Figure \ref{fig:sfr-mass} does see some changes in the shape of the distribution of the control. However, we find the red sequence consistent throughout each stage, and it does not reduce like in the interacting sample. As shown by the surrounding bar chars of the scatter distributions, the control SFR distribution remains largely consistent through each stage.

However, this evolution could also be due to many other factors rather due to interaction stage. For instance, if the stellar mass distribution of our sources evolves we could simply be selecting higher stellar mass systems as we increase stage. This would have the result of systems in the merging stage having, on average, higher SFRs than those in the separated stage and appearing like we had evolution in the star forming population with stage. Another effect that could cause this relation to appear would be our selection was highly dependent on galactic environment. However, we have already explored this possibility in Section \ref{data:environ} and found no bias in environment between the stages.

We can quantify the similarity of the stellar mass distributions, and then the SFR distributions, using the KS-test. First, we create weighted distributions of stellar mass. The weighting scheme balances the distributions such that each bin could be assumed to have the same number of sources within it. Therefore, any bins with fewer than a certain number of sources will be weighted up while those bins with more will be weighted down. These weights applied to the stellar mass distribution are then applied to the SFR distribution as to control for stellar mass in this distribution.

\begin{figure*}
    \centering
    \includegraphics[width = \textwidth]{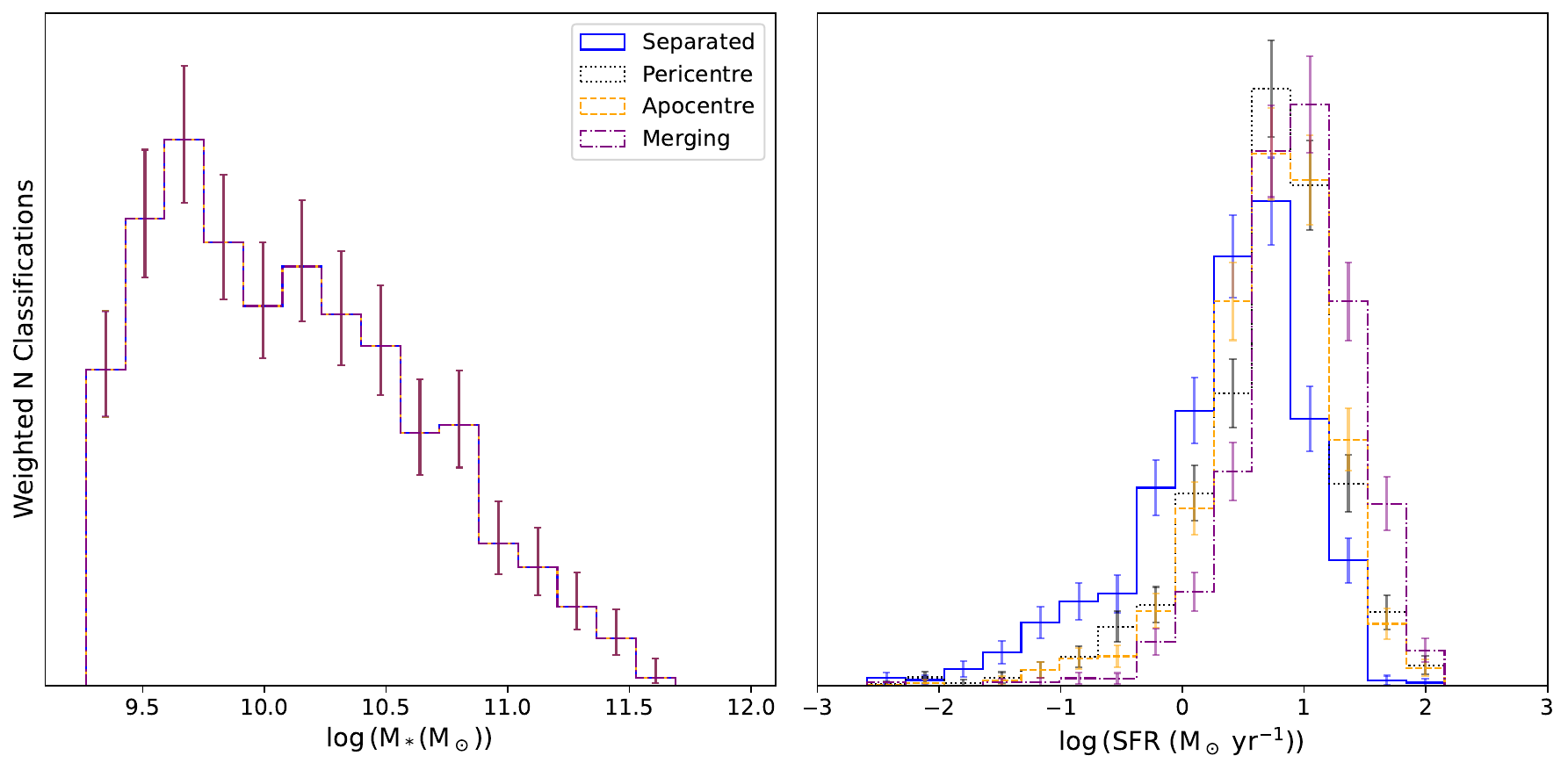}
    \caption[The weighted stellar mass distribution across the four stages.]{The distributions of stellar mass and star formation with stage weighted by stellar mass counts. \textit{Left}: The weighted stellar mass distribution across the four stages. Each bin is weighted based on the counts in the smallest sub-sample in stage: the merging stage of the interaction. The weighting is designed to control for stellar mass across the samples, for cleaner comparison of other properties. \textit{Right}: SFR distribution weighted by stellar mass across each stage. This weighting is based on our sample of merging stage.}
    \label{fig:weighted-mass-sfr}
\end{figure*}

The left panel of Figure \ref{fig:weighted-mass-sfr} shows the weighted stellar mass distributions through the four different stages. We chain the KS-test through each distribution and calculate KS-scores and the $p$-values. Table \ref{tab:ks-scores} shows the resultant KS test scores. Larger KS scores correspond to smaller $p$-values. The $p$-value represents the probability that each distribution is drawn from the same parent distribution. For each stellar mass distribution, we find that the KS score is close to zero and the $p$-value of the test is $\approx1$. Therefore, the distributions in stellar mass though each stage are consistent with being drawn from the same parent sample. 

\begin{table}
    \centering
    \resizebox{\columnwidth}{!}{\begin{tabular}{|c|c|c|c|c|}
        \hline 
        \multicolumn{5}{|c|}{KS Scores for Stellar Mass (KS Scores for SFR)} \\
        \hline
        Stage & Separated & Pericentre & Apocentre & Merging \\
        Separated & 0 (0) & 0.019 (0.242) & 0.016 (0.233) & 0.015 (0.418) \\
        Pericentre & - & 0 (0) & 0.014 (0.034) & 0.014 (0.187) \\
        Apocentre & - & - & 0 (0) & 0.003 (0.195) \\
        Merging & - & - & - & 0 (0) \\
        \hline
    \end{tabular}}
    \caption{The KS scores resulting from applying the weighted two sample KS-test to each stage. First is the scores between the mass distributions and in brackets is the scores between the SFR distributions.}
    \label{tab:ks-scores}
\end{table}

The right panel of Figure \ref{fig:weighted-mass-sfr} shows the SFR distributions while being weighted by stellar mass. The resultant KS scores are shown in brackets in Table \ref{tab:ks-scores}. The resultant scores are much larger than when comparing the stellar mass distributions. We find when comparing the separated and pericentre, separated and apocentre, separated and merging, pericentre and merging and apocentre and merging stages, the $p$-values are $\ll$0.05 ($\ll2\sigma$). This allows us to reject the null hypothesis for these distributions and assume they are from different parent samples. However, for comparing pericentre and apocentre stages, the $p$-value$=0.74$. Thus, while these distributions are likely to be not identical, their parent samples are likely similar. For the same stellar mass distribution through each stage of interaction, the star formation distribution changes from separated to pericentre stage and from apocentre to merging stage, while remaining similar from the pericentre to apocentre stage. The errors on these distributions are calculated using the methodology of \citet{2011PASA...28..128C}. In this approach, we assume that the error about our measurements is simply a measure of the limited samples we have of the underlying beta distribution of the full population. Thus, we can measure the confidence intervals around our measurement by using the beta distribution. We calculate the 68.3\% confidence interval about our found value, and use the distance between these values as our upper and lower errors.

Putting this result into the context of the dynamical timescale of an interaction, it shows there are distinct points at which the SFR changes in these systems. The first is when the interacting system moves from the close pair stage to morphologically disturbing each other in a close flyby. This change then persists through to the apocentre stage - where the galaxies remain highly disturbed but are no longer overlapping with their secondary. The SFR distribution remains approximately the same between these two, meaning the forces that drive and affect star formation remain equivalent between these two stages. Finally, the SFR changes again when we approach the merging or post-merger stage of the interaction. We can also say that this change is likely an enhancement in the SFR of the galaxies through the interaction due disturbance and movement of the gas within each galaxy. We fully explore the reasoning for this in Section \ref{disc:int_stage}.

We further examine this result by controlling for redshift and classifying our galaxies based on the star formation. We have shown that the red sequence is significantly reduced across each stage, but it is difficult to ascertain the change in the galaxies within the blue cloud. We adopt the stellar mass- and redshift-dependent star-forming classifications of \citet{2019MNRAS.484.4360A} for all galaxies in our system. They define the star forming main sequence (SFMS) as:

\begin{equation}
    \log \text{SFR}_{\text{SFMS}}\left(z\right) = -7.6 + 0.76\log\frac{M_{*}}{M_{\odot}} + 2.95\log\left(1+z\right).
\end{equation}

The measured scatter about the SFMS then leads to a classification of how intensively star formation is occurring in the galaxy. These classifications are starburst (SB), main sequence (MS), sub-main sequence (S-MS), high quiescent (Q(High)) and low quiescent (Q(Low)). The full criteria for each classification is given in in \citet{2019MNRAS.484.4360A}). In total, we find 620 SB galaxies, 1,979 MS galaxies, 540 S-MS and 239 quiescent galaxies (of which 149 are Q(High) and 90 Q(Low)).

\begin{figure}
\centering
\includegraphics[width=\columnwidth]{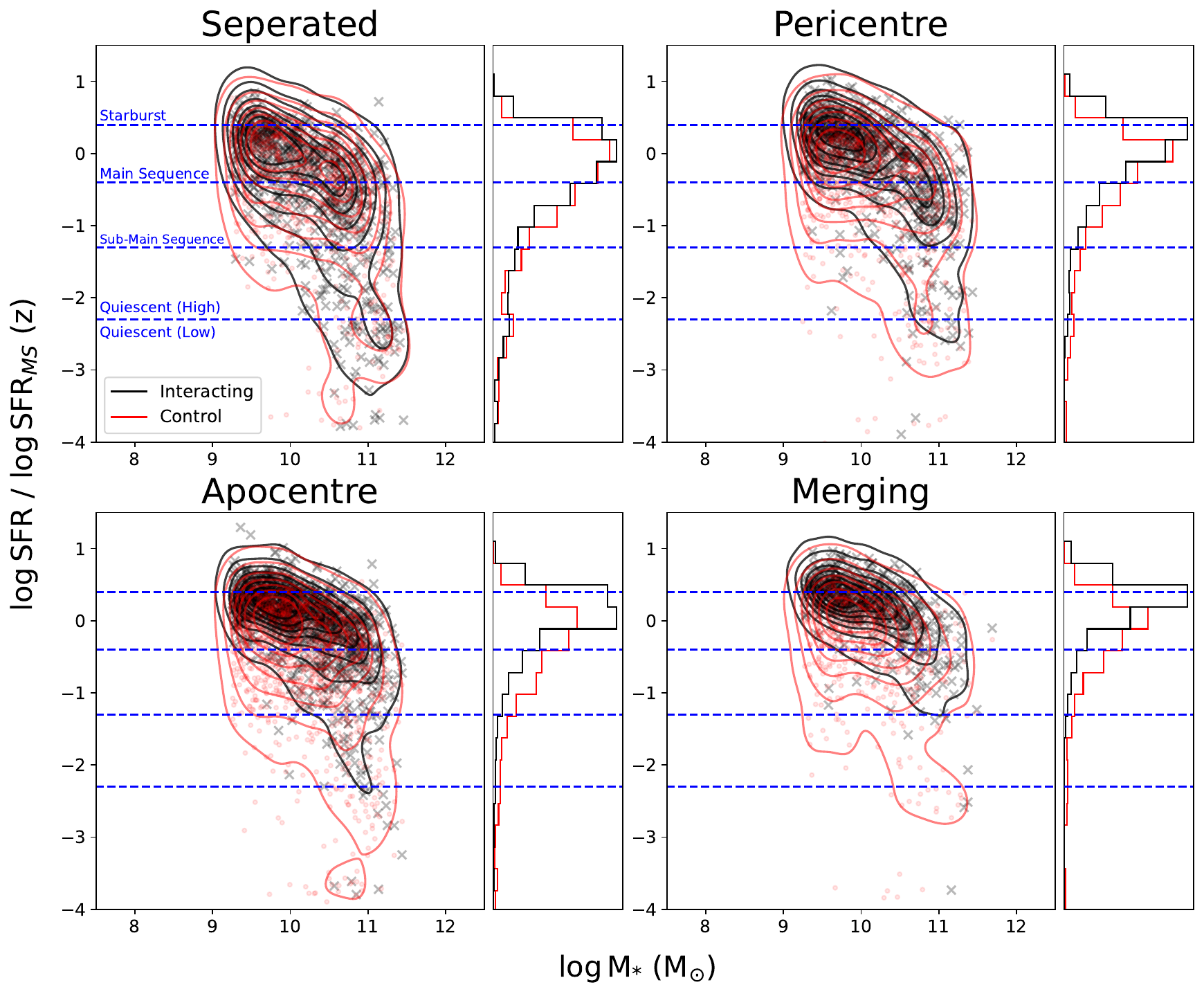}
\caption[Stellar mass vs the ratio of measured SFR to the expected SFR if the galaxy was on the SFMS.]{Stellar mass vs the ratio of measured SFR to the expected SFR if the galaxy was on the SFMS. The black points and contours are the distribution of our interacting sample while the controls are shown in red. The blue dotted lines show the cutoffs for different galaxy classifications based on their SFR, with each cut off being defined by the text in blue. The histograms beside each plot show the change in counts. We find that through interaction stage, the quiescent galaxy population significantly reduces while the starburst population rapidly increases while there is little change in the control. As these cutoffs are also dependent on redshift, we find that this evolution in SFR with interaction stage is independent of redshift.}
\label{fig:sfr-clsf}
\end{figure}

\begin{figure}
\centering
\includegraphics[width=\columnwidth]{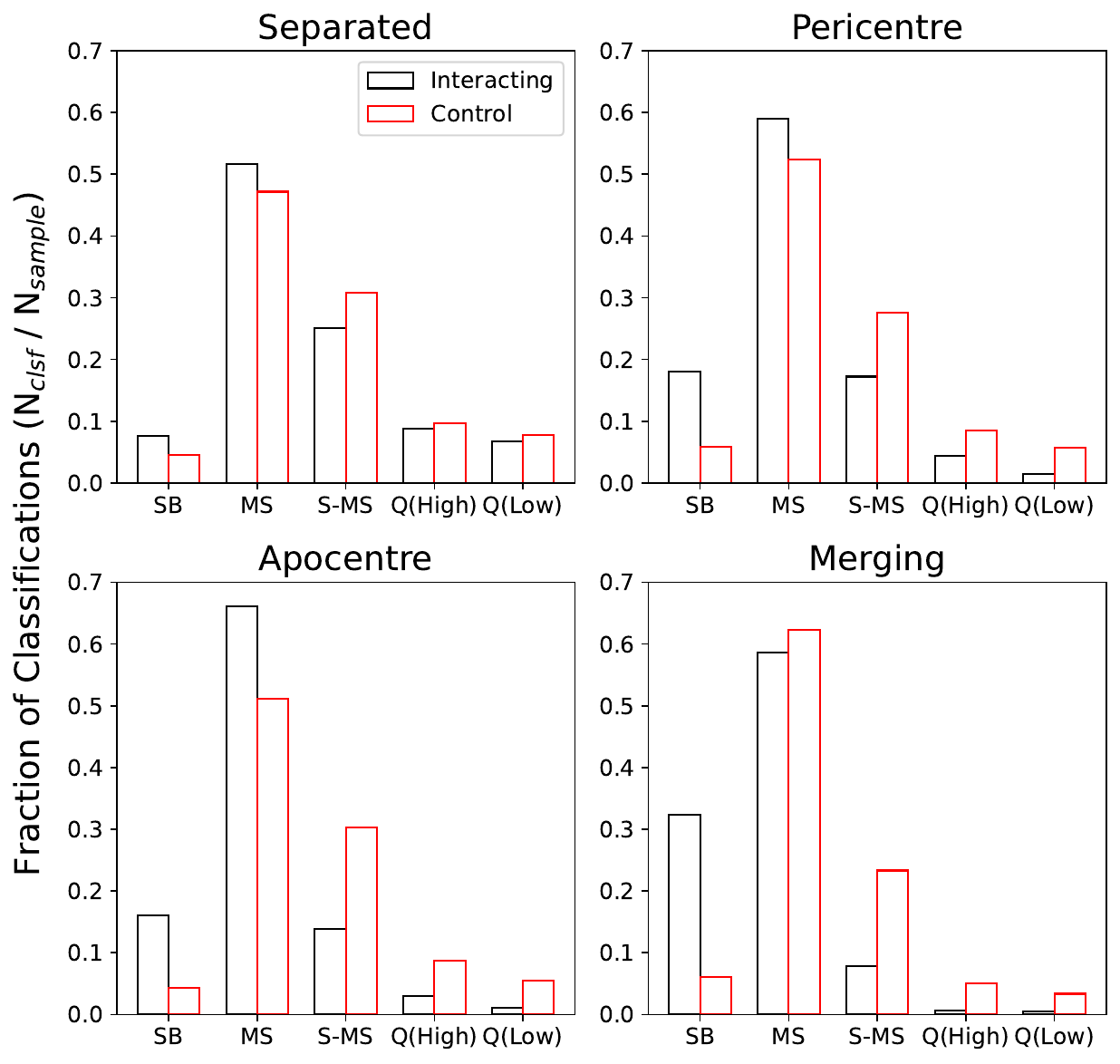}
\caption[The change in fraction of different galaxy classifications from the fraction of SFR to the expected SFR on the SFMS.]{The change in fraction of different galaxy star formation classifications. In black we show the interacting galaxy fraction while red shows the control galaxy sample. In the interacting sample, the galaxies on the SFMS remain dominant through each sample across interaction stage, there is significant change in the starburst and quiescent populations. The starburst galaxy population moves from being roughly half the size of the quiescent population in the separated stage to completely dominating it in the merging stage. This is occurring while the quiescent population is significantly reduced to almost non-existence in the merging stage.}
\label{fig:sfr-clsf-bar}
\end{figure}

Figure \ref{fig:sfr-clsf} shows the ratio between the expected SFMS SFR and the measured SFR in the C20 catalogue. In black of the interacting sample and in red of the control sample. This clearly shows a large increase in galaxies classified as starburst from the separated to merging stages and a large reduction in the number of quenched systems. Figure \ref{fig:sfr-clsf-bar} shows the change in fraction of the different galaxy classifications through stage, reflecting the results found in Figure \ref{fig:sfr-mass} with the same colour scheme. Initially, in the separated stage, we find that the majority of our galaxies lie on the SFMS or just below it. There also exists a small population of galaxies which are classified as starburst with a population of quiescent galaxies that is roughly double the starburst fraction. As we move through the interaction stage, we see that the quiescent galaxy fraction gradually decreases to the point of almost non-existence in the merging stage galaxies. The inverse is true in our starburst fraction. We find this almost quadruples over the course of the different stages of interaction. The fraction of galaxies on the SFMS remains dominant throughout, however, we do find the fraction of sub-MS galaxies significantly reduced. The SFR of these galaxies is increasing with interaction stage (though, not in the pericentre to apocentre stage). The population of sub-MS and quiescent galaxies increase their SFR and join the SFMS. But, many galaxies from the SFMS are moved upwards and into a starbursting phase. This shows a general trend of increased star formation as the interaction progresses.

It is important to note the parameter space that we are searching in this sample. We are probing interactions between galaxies of high stellar mass, where the resultant tidal features that form would be classifiable in an image. The majority of our sample is minor interactions where one of the two systems is very highly perturbed by the interaction. In this parameter space, we would expect large increases in the SFRs. However, we do not see a significant increase in the starbursting population until we find the two galaxies begin to actually coalesce. Thus, the effect of the interaction itself may be to only enhance the SFR, while at coalescence we find the dramatic starburst. This can be argued from our results of evolution of a galaxy's SFR with interaction stage.

\subsection{Projected Separation and Star Formation Enhancement}
\noindent In this work, we have used a staging system to classify the point in the dynamical time an interaction is occupying. More often in the literature, the projected separation between the two systems is used. We directly investigate the relation between the SFE and the projected separation using our confirmed sub-sample of galaxy pairs. This sample is significantly smaller than our non-pair sample: containing 556 pairs or 1,112 galaxies. Upon sub-dividing this into different stages we find 245 separated galaxies, 93 pericentre galaxies and 224 apocentre galaxies. Only 4 merging galaxies were in our confirmed galaxy pair sub-sample, and therefore we do not attempt to make inferences about this population. 

To measure the SFE of our galaxy pairs, we directly compare to the stellar mass- and redshift-matched control sample that was also created with this sub-sample and defined in Section \ref{sec:sec-ident}. We separate our galaxy pairs into different bins based on the projected separation between them. We find that the bulk of our sample has a projected separation between the two galaxies of $\leq$50kpc. Therefore, we sample from this region of the parameter space with high precision and smaller bin widths before we increase the bin widths at larger projected separations. We define a cutoff that each bin must contain at least 10 counts to be included in this plot and, therefore, by changing the bin widths with projected separation we are able to maintain some level of statistical robustness. We define our bins as [0.5, 10, 20, 50, 100, 125] kpc.

We create two projected separation distributions: one of interacting galaxies and the other of our control galaxies. We then take the average of each bin. By taking the average of each bin, we find the total excess of the SFR caused by interaction alone and then compare it to what we would expect from non-interacting galaxies in the same bin. Note, that the control galaxy relation to the projected separation is meaningless as they are not paired. The control galaxies are simply stellar mass and redshift matched to each interacting galaxy and used as a reference for what SFR we would expect of it had it not been interacting. Finally, we divide the averaged interacting SFR distribution by the averaged control SFR distribution. This provides us with a measure of the excess star formation due to interaction, and the SFE when compared to the isolated population.

\begin{figure*}
\centering
\includegraphics[width=\textwidth]{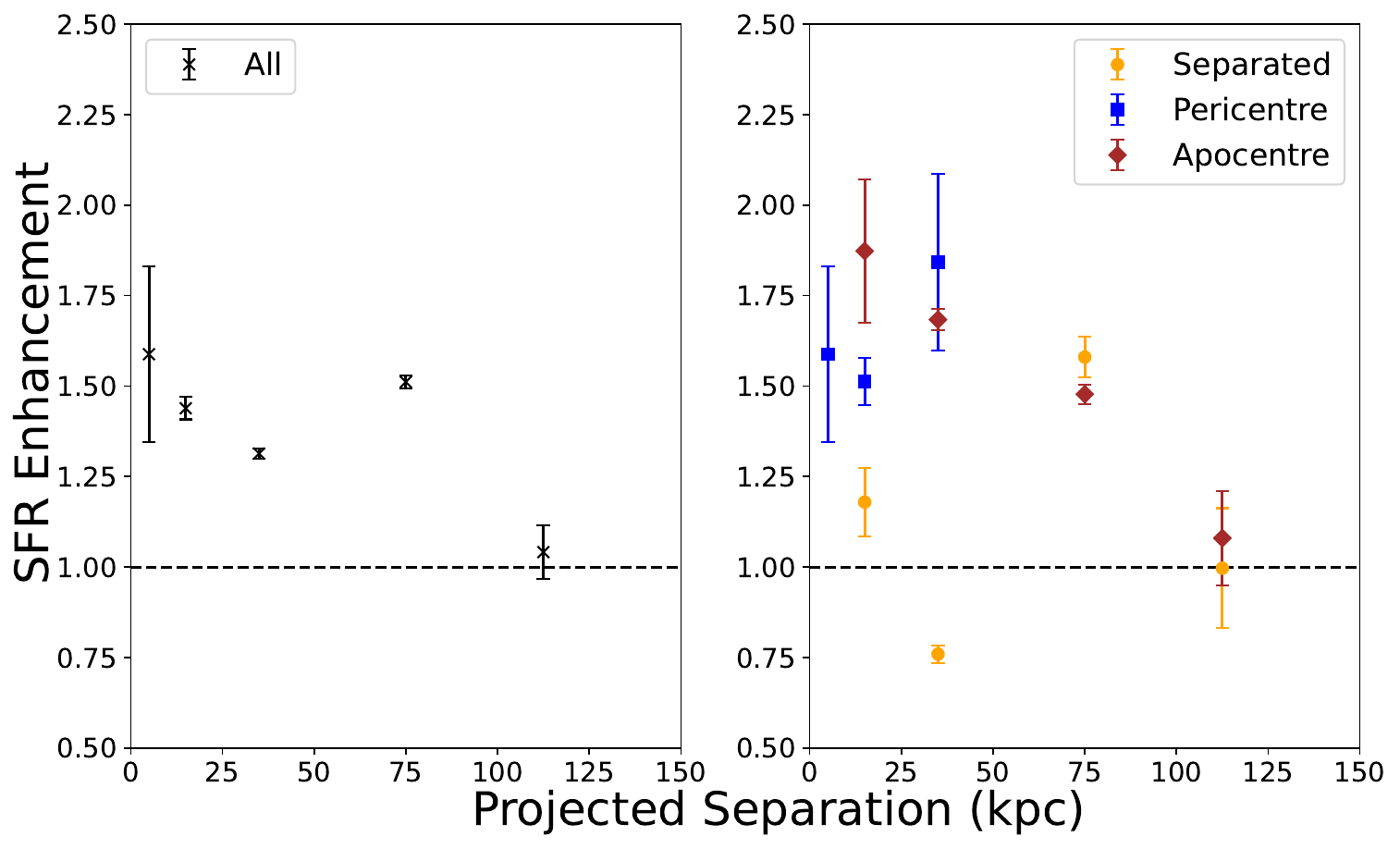}
\caption[The projected separation against the SFE in average star formation at different bins of projected separation.]{The projected separation against the SFE in average star formation at different bins of projected separation. Each bin must contain at least 10 counts to be considered. As the bulk of our galaxy pair sample is at low projected separation, we heavily sample from this region of the parameter space. The bins are: [0.5, 10, 20, 50, 100, 125] kpc. As we move to higher projected separation, the bins increase in width to maintain statistical significance in our sample. \textit{Left}: The SFE found in the full galaxy pair sample. As expected, we see a gradually decreasing enhancement. \textit{Right}: As left but broken up into different stages of the interaction. Yellow circles are the separated stage, blue boxes the pericentre stage and brown diamonds the apocentre stage. We limited our investigation to the separated, pericentre and apocentre stages as only three galaxy pairs were identified in the merging stage. We find a generally decreasing star formation enhancement with projected separation but very different individual behaviour dependent on the stage classification.}
\label{fig:sfr-enhancement-sep}
\end{figure*}

Figure \ref{fig:sfr-enhancement-sep} shows star formation enhancement between our interacting and control binned SFRs and the projected separations between each galaxy pair. On the left of the plot, we have the distribution for the galaxy pair sample, without taking account of stage. The errors on our measurements are following the methodology of \citet{2011PASA...28..128C} and briefly described in Section \ref{results:int_stage}. The dashed black line represents no enhancement in SFR, as the average SFR in the interacting galaxy sample would be equal to the average SFR in the control sample. We find at very small projected separation a SFR enhancement of 1.74 which gradually decreases with projected separation down to approximately 1 at 112.5kpc. 

On the right, we break the galaxy pair sample into different stages and plot the resultant enhancement in star formation. In the separated stage, we find that the star formation is very weakly enhanced, and with no overall structure or trend. The enhancement moves around 1 through the distribution, with some enhancement being a result of low number counts in the bin. This is the stage with the lowest enhancement across projected separation.

In the pericentre stage, we see consistent enhancement with projected separation. It is also much larger than the separated stage, with it being approximately 1.6. In fact, the measured enhancement increases across the 50kpc of projected separation. However, accounting for the errors on these measurements and the declining counts in each bin, it is likely that the enhancement remains constant as the projected separation increases. We see a very large enhancement in the apocentre stage which overlaps with the pericentre stage. The enhancement in apocentre stage galaxies then rapidly declines as we move to higher projected separations. After 150kpc, our galaxy pair sample does not have the counts to make robust estimates of the SFE.

\section{Nuclear Activity with Interaction Stage}\label{results:AGN_stage}
\noindent We now use our AGN described in Section \ref{sec:agn-clsf}. As stated previously, we find 802 matches of AGN and SFGs. A breakdown of the classifications is shown in \ref{tab:agn-sfg-breakdown}. We assume that any source in our catalogue that has not been included in the VLA 3GHz or Chandra catalogues is not an AGN. We remove sources that would lie outwith the footprint of the VLA 3GHz catalogue, and do not assume that these are undetected SFGs.

\begin{table}
    \centering
    \begin{tabular}{|c|c|c|c|c|}
        \hline
        & Separated & Pericentre & Apocentre & Merging \\
        \hline
        SFG & 135 & 212 & 337 & 199 \\
        AGN & 103 & 117 & 163 & 95 \\
        \hline
    \end{tabular}
    \caption{Breakdown in number of classified AGN and SFGs per stage.}
    \label{tab:agn-sfg-breakdown}
\end{table}

\begin{figure*}
    \centering
    \includegraphics[width=0.95\textwidth]{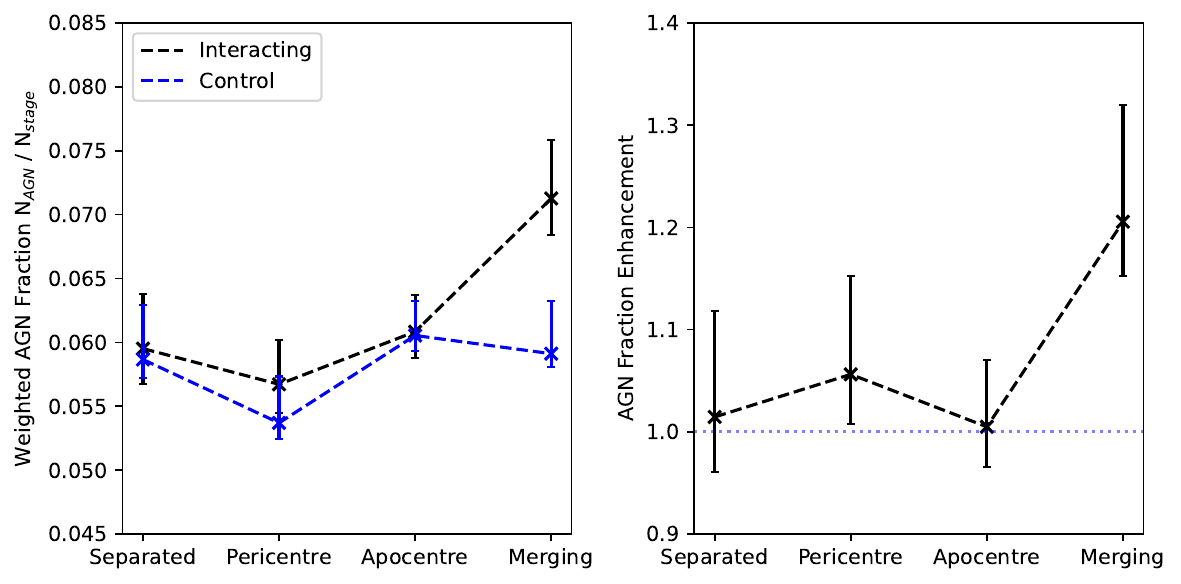}
    \caption[The change in AGN fraction with stage.]{The change in AGN fraction with stage. We have defined the AGN fraction as the ratio of galaxies hosting an AGN to the total number of galaxies in each stage. The contributions of individual galaxies have been weighted by the stellar masses as in the previous SFR distribution plots. \textit{Left}: the weighted fractions of our interacting galaxy sample (black) and of the matched control sample (blue). \textit{Right}: the ratio of the interaction AGN fraction to the control AGN fraction. This shows the potential enhancement in the fraction due to the interaction. The dotted line indicates a ratio of 1: no enhancement. Errors on each fraction are found as the confidence intervals defined via the beta distribution. We find slight enhancement in the AGN fraction at the pericentre stage, but major enhancement at the merging stage.}
    \label{fig:agn-frac-time}
\end{figure*}

Figure \ref{fig:agn-frac-time} shows the changing AGN fraction with interaction stage. The AGN fraction has been calculated as the ratio of the number of galaxies containing an AGN to the total number of galaxies classified in each stage. The contribution of each galaxy to this ratio has been weighted by the stellar mass as when applying the KS-test. This normalises the counts, and assumes that we have equivalent numbers of galaxies in each stellar mass bin for each distribution. On the left in black we show the measured fraction of AGN in our interaction sample, while the blue shows the measured fraction of AGN in our control sample. The errors are measured using the beta distribution, and are measured as the 68.3\% confidence interval. On the right, we show the AGN 'enhancement' - the ratio of AGN in the interacting sample to the control sample.

We find that the AGN fraction generally remains at the fraction of the control sample through the separated to apocentre stages. This oscillates around an AGN fraction of $\sim0.06$. We find slight enhancement at the pericentre stage but this remains approximate to the control sample. In the merging stage we measure a large increase in the AGN fraction from $\approx0.06$ to $\approx0.07$, an enhancement of 1.2.

\section{Discussion}\label{discussion}
\subsection{Interaction Stage and Projected Separation}\label{disc:int_stage}
\noindent Finding evolution in galactic parameters, such as SFR, with interaction is not new in the field. Multiple works have found increases in SFE and SFR as a function of projected separation \citep{2000ApJ...530..660B, 2008AJ....135.1877E, 2013MNRAS.433L..59P}. Projected separation is often used as a proxy for the different epochs of dynamical time of the interaction. It is likely that, when pairs of galaxies are closer together, they are closer to coalescence in the dynamical time. However, this fails to capture the larger complexity of interaction. Without morphological consideration, it is unclear whether galaxies at small projected separations are actually at the closest point of flying by each other or about to coalesce. This can be accounted for by considering the morphological signatures in the system.

We find that throughout the different stages of interaction the SFR is increasing. Observations of interacting galaxies at various stages have found increased gas inflows into the nuclear regions of the galaxies which lead to enhancement in the SFR at the galactic core over their outer reaches \citep{2015A&A...579A..45B}. This is often confirmed with deep observations of individual systems, which capture snapshots of different parts of the dynamical timescale of the interaction \citep{2022MNRAS.514.2769K}. Our study utilises snapshots of the dynamical time of interaction in attempt to build a full picture of the change. We have found that this process increases the SFR of galaxies and continuous increases in the starburst fraction from the pericentre of interaction. Multiple simulation works, which have the ability to model the entire dynamical history, show that this is to be expected \citep{2007A&A...468...61D, 2013MNRAS.430.1901H, 2015MNRAS.452.2984K, 2021MNRAS.503.3113M}. Often, these works show an initial dramatic increase in the SFR of interacting systems followed by an exponential decline through the dynamical time before increasing dramatically again at either a second passage or coalescence. \citet{2015MNRAS.448.1107M} is a direct example of this star formation history through the interaction.

Our results differ here as we do not find an exponential decrease in the SFR as our classification changes from pericentre to apocentre, but a gradual increase through the interaction. Other observational works do support a continued increase in SFE in galaxies to projected separations of out to 80kpc - well into our defined apocentre stage systems \citep[for further examples, see][]{2008MNRAS.385.1903L, 2012MNRAS.426..549S}. However, it is important to note that our apocentre classification is a `wide net' that captures many systems that may be very soon after the initial passage in the dynamical timescale (recall Figure \ref{fig:illustration}). The criteria defining pericentre and apocentre stages are simply that the galaxies must be no longer connected or overlapping morphologically with tidal features. They must only be distinct and separate galaxies. Therefore, our found large enhancement in the apocentre stage may be from interacting systems which are only just out of the pericentre passage and not enough time has passed for the rapid decline in SFR to begin. 

Nonetheless, we still find a disappearance of the red sequence in the apocentre systems which may come as unexpected when compared to simulations. It is important to note, however, that this is not the same as saying that a large proportion of the apocentre systems are classified as starbursting galaxies. While simulations approximate an initial large starburst before rapid decline, we find the interacting galaxy population is pushed from being quiescent / sub forming main sequence to being on the star forming main sequence. Therefore, it may be more likely that the impact of interaction on star formation is not to suddenly cause rapid star formation in the aforementioned starburst before declining, but rather to gradually increase a galaxy's SFR up and into the blue cloud of galaxies. This could be supported by the lack of rapidly quenched post-interaction galaxies found in both observations \citep{2017ApJ...845..145W} and simulations \citep{2020MNRAS.493.3716H, 2021MNRAS.504.1888Q}. Thus, the impact of interaction on star formation may not be a sudden increase in star formation which exhausts a galaxies gas reservoir and leads to quenching. Instead, the impact may be a small increase in a galaxies SFR over a long period of time leading them to quench slowly. 

This idea can be further explored by considering the large increase in starbursting galaxies we find when changing from the pericentre / apocentre stage to the merging galaxies. The merging stage represents, in our sample, galaxies that are undergoing the final coalescence of the two systems involved. We find that the point of final coalescence leads to a large increase in the fraction of starbursting galaxies as well as the almost complete disappearance of the quiescent and sub-main sequence fraction of galaxies. We can conclude that during coalescence galaxies undergo a starburst and enhancement in star formation which will quickly lead to their quenching. This is also supported by works such as \citet{2022MNRAS.517L..92E}, which find that galaxies post-coalescence are 30-60 times more likely than control galaxies to have rapidly shut down star formation.

For merging galaxies if gas is present within them, star formation will increase and change our classification of the galaxy. We will observe a sudden increase in the SFRs of these systems followed by a rapid quenching as the gas is used up entirely. This differs from galaxies that move into the apocentre stage and do not merge. These apocentre stage systems will then slowly lose their enhancement, returning to star forming at their expected rate.

There has been limited discussion in the literature on whether interaction leads to significant enhancement \citep{2007A&A...468...61D, 2020MNRAS.494.4969P} or not \citep{2015MNRAS.454.1742K, 2019A&A...631A..51P}. The only part of the pre-merger interaction which causes the enhancement, and is the decisive point, is the pericentre stage. After this, if the galaxies then advance to the apocentre stage and escape, we will see a gradual decline in the SFE with the apocentre stage galaxy only having a minor increase in its star formation classification. Whereas if the galaxies move into the merging stage of the interaction and coalesce we see the results of a major starburst and the complete using and of all the gas in the systems. Leading to a large fraction of quenched post-merger galaxies, but only after coalescence. 

\subsection{Interaction Stage and AGN Fraction}

\noindent The evolution of the AGN fraction in interacting galaxies is similar to that of the evolution of star formation. It has often been found with projected separation the AGN fraction increases. There are multiple observational works that show this \citep{2007MNRAS.375.1017A, 2013MNRAS.435.3627E, 2020ApJ...904..107S} as well as works on cosmological simulations which support these conclusions \citep{2023MNRAS.519.4966B}. In simulations, the increased likelihood of AGN activation comes from the sudden increase in gas density in the galactic core which leads to increased black hole feeding, growth and nuclear activity. Observations of interacting galaxies often contain examples of dual AGN and individual examples \citep[e.g.][]{2017MNRAS.470L..49E, 2021ApJ...923...36S} or investigate the increase in the AGN fraction in only the merger / post-merger stage \citep{2020A&A...637A..94G}.

It has been found, however, that there is also an enhancement in the fraction of AGN in post-starburst galaxies \citep[PSBs][]{2023MNRAS.519.6149B} as well as a likely connection between the formation of PSBs and the merging of systems \citep{2022MNRAS.516.4354W}. The resultant use, removal, or introduction of ionisation and turbulence of the gas then leads to a rapid decline in the star formation in the galaxy \citep{2018MNRAS.478.3447E}. However, the role of the AGN in this process, and specifically in the formation of post starburst quenched galaxies, remains unclear - whether it is an active part of the quenching process in the PSB or is a result of other effects \citep{2019MNRAS.484.2447D, 2023MNRAS.523..720L}. Our work is evidence that the AGN switch on is during the merging process, which could then act as to remove gas and excite the remaining gas to the point where star formation is stopped. 


The link between a higher AGN fraction in a merger is also supported with other works focused on AGN fraction with projected separation \citep{2008AJ....135.1877E, 2011MNRAS.418.2043E, 2018PASJ...70S..37G, 2021MNRAS.504.4389D, 2023ApJ...942..107S}. Such works have found the AGN fraction increases during the points in the dynamical time when the inner parts of the galaxy are majorly disturbed, and not by the simple movement of gas and dust into the core during the two galaxies passing each other - we would have, therefore, found enhancement at the pericentre stage. This implies that a further mechanism is required for nuclear activation than only more gas being present \citep{2007MNRAS.375.1017A}. There must also be further disruption to the gas to cause the infall to begin, and the black hole to begin growing. This approach is supported by simulations \citep{2020MNRAS.494.5713M, 2023MNRAS.519.4966B}, which find the AGN enhancements at the merging stage rather than others, but not that a merger will ensure that nuclear activation occurs. They ascribe the differences to the properties of the merger; e.g. the mass ratio, the gas content.

This matches both what we find and what observations found previously of increased gas densities in nuclear regions and cores, and that the merging and post-merging stage is when this really occurs in earnest. However, to more fully study this, we would need a larger sample of confirmed AGN and star forming galaxies from existing photometry or catalogues to make more decisive conclusions based on stage. To fully identify the processes leading to nuclear activation in a merger, or not, we would need deep integral field spectroscopy data in a pure sample of merging galaxies which captured the gas kinematics of the core.

\section{Conclusions}\label{conclusion}
\noindent In this work, we investigate the evolution of galactic parameters and processes in interacting galaxies with interaction stage. We use the interacting galaxy catalogue created in OR23 and match it with the COSMOS survey for ancillary data. This gives us a flux limited sample of 4,181 interacting galaxies of which 845 have a confirmed secondary from available photometric redshift data. We apply a stellar mass limit to our sample, reducing it to 3,162 galaxies with 556 pairs. We use visual morphology as well as angular separation to split our sample into four distinct stages: (1) close pair, (2) morphologically disturbed and overlapping, (3) morphologically disturbed and distinct, (4) merging. Each stage is designed to capture a different part of the dynamical timescale of the interaction. We match these samples with existing catalogues of environment and AGN.

We first split our sample of galaxies into their different stages and investigate their evolution with stellar mass and SFR. First, we check for any biases in galactic environment with stage. Environment is known to have direct effects on the observed star formation of galaxies from a range of processes including harassment, strangulation and ram pressure stripping - particularly if many of our interacting galaxies are found in clusters. We find no bias of environment with stage, and have a consistent distribution of our sources in the field, filaments or in clusters. We also discuss other potential sources of contamination such as incorrect redshifts, similar morphologies and clumpy galaxies.

With these checks completed, we apply KS-tests to the stellar mass and SFR dstributions of our sample. We find the stellar mass distribution does not change, while the SFR distributions evolve with interaction stage. We find that the fraction of starburst galaxies in our sample change from 7\% in our close pair stage, to 32\% in our merging stage. The red sequence declines over the same interaction stages from 21\% to $<1\%$. This change in the SFR distribution from the separated to merging stages is found in the red sequence of galaxies reducing to the point of disappearance. This is further confirmed by sub-classifying each sampled stage into starbursting, main sequence and quiescent galaxies. We find that as the galaxies move from the separated to pericentre stages the fraction of starbursting galaxies increases while the quiescent galaxy fraction reduces. In the merging stage, the starbursting fraction increases dramatically while almost no quiescent galaxies exist in the sample. This implies that the mechanisms responsible for enhancement in star formation in interacting galaxies is dominant from separated to pericentre stages and in the final coalescence of the system. We find that, for all of our galaxies, some enhancement in star formation is observed. 

To further investigate this change in enhancement, we investigate our sub-sample of galaxy pairs and compare it to a stellar mass and redshift matched control sample. We bin our projected separations and measure the ratio between the average interacting and control SFR in each bin. We find, across the whole subsample, a general increase in SFE as we move to smaller projected separation. The highest found enhancement is below $10~\mathrm{kpc}$. However, when broken down into their constituent stages, we find different behaviour in the SFE. This is best seen in the pericentre stage enhancement, which does not seem to change with projected separation while remaining enhanced. There is a dramatic decrease in the apocentre stage enhancement from 1.8 at ${\sim} 10\mathrm{kpc}$ down to 1.0 at ${\sim} 125\mathrm{kpc}$. This shows that just using projected separation as a proxy for stage will leave out crucial information to the underlying causes and mechanisms fueling increases in the SFE. To confirm that the effects observed here are not related to biases in the environment, we investigate its relation to our staged sample. We find that the environment is consistent between all stages.

Finally, we investigate the change in AGN activity across our whole sample. We find that, on average, the AGN fraction remains constant with stage until the point of coalescence. We find a slight enhancement in the AGN fraction compared to our control sample of $1.19\pm0.04$. This is inline with other works, both on galaxies undergoing a merger but also in studies of PSB galaxies. We find a slight enhancement at the pericentre stage of $1.04\pm0.04$, however, this is also consistent with no enhancement. The majority of our AGN sample are found using radio emission, rather than X-ray emission, which has been found to affect when in an interaction an AGN is detected.

While the results with projected separation are not unexpected, those with the different stages are. We have shown the use of projected separation of interacting galaxies as a proxy for the stage of the interaction may remove crucial information. We find differing behaviour in the SFR of interacting galaxies based on stage which are at in the dynamical timescale of the interaction. We find the beginning of an enhancement in star formation occurs in the pericentre stage. Then, through to the apocentre stage, the enhancement remains although with no further increase. There is then another increase in the SFR of the interacting population in the merging stage - represented by an increase in the fraction of starburst galaxies in the sample.

Our work shows the importance of considering morphological stage when considering interaction, and that there is a fine interplay between underlying processes and dynamical timescale of an interaction. We require larger samples of correctly staged galaxies to further understand and exploit what these relations are, and how to best investigate when and where star formation and its enhancement occurs. This is also particularly true of the relation between AGN and the interaction stage. We require numerical models to better identify the point in the dynamical timescale an interacting system is. This would lead to more quantative constraints on the relations we have explored in this work.

\section*{Acknowledgements}
The authors thank the anonymous referee for their comments on this paper. Due to their time and effort, this manuscript became a much stronger work than without their input.

This project was conducted as part of DORs PhD program supported by the UK Science and Technology Facilities Council (STFC) under grant reference ST/T506205/1. It was finalised while conducting a postdoctoral research position at the Centro de Astrobiologia (CAB), INTA-CSIC. This was under grant number 2012/212. BDS acknowledges support through a UK Research and Innovation Future Leaders Fellowship [grant number MR/T044136/1]. ILG has received the support from the Czech Science Foundation Junior Star grant no. GM24-10599M. The Cosmic Dawn Centre (DAWN) is funded by the Danish National Research Foundation (DNRF) under grant No. 140.

This research made use of many open-source Python packages and scientific computing systems. These included \texttt{Matplotlib} \cite{matplotlib}, \texttt{Pandas} \citep{pandas},and \texttt{numpy} \citep{numpy}. This work also extensively used the community-driven Python package \texttt{Astropy} \citep{astropy:2018}.

The data used in this work came primarily from the COSMOS survey. This is a large dataset from a host of other observatories under the digital object identifier (DOI) 10.26131/IRSA178. Under this identifier, we have made use of the Classic and Farmer COSMOS2020 catalogues. These are based on observations collected at the European Southern Observatory under ESO programme ID 179.A-2005 and on data products produced by CALET and the Cambridge Astronomy Survey Unit on behalf of the UltraVISTA consortium. The original classifications of interacting galaxies were made using data for the Advanced Camera for Surveys aboard the \textit{Hubble Space Telescope}.

Finally, DOR would like to again thank the students who made this work possible: OMC, MR, ERW, DW and SGVP. Their initial classifications of the interacting stages paved the way for this work.

\section*{Data Availability}
\noindent The interacting galaxy catalogue used in this work is available from \citet{2023ApJ...948...40O}.

Other catalogues that have been used in this publication are:

\begin{itemize}
    \item COSMOS: \newline
    DOI: \href{https://www.ipac.caltech.edu/doi/irsa/10.26131/IRSA178}{10.26131/IRSA178}. This survey was described in \citet{2007ApJS..172....1S}. The COSMOS2020 catalogue is described in \citet{2022ApJS..258...11W}.

    \item MPA-JHU: \newline
    \href{https://wwwmpa.mpa-garching.mpg.de/SDSS/DR7/}{https://wwwmpa.mpa-garching.mpg.de/SDSS/DR7/}
\end{itemize}

\noindent Each of these catalogues are publicaly available.


\bibliographystyle{mnras}




\bsp	
\label{lastpage}
\end{document}